\documentclass[a4paper,fleqn,usenatbib]{mnras}

\usepackage{newtxtext,newtxmath}

\usepackage[T1]{fontenc}
\usepackage{ae,aecompl}

\usepackage{float}
\usepackage{graphicx}	
\usepackage{amsmath}	
\usepackage{amssymb}	
\usepackage{bm}
\usepackage{color,hyperref}

\newcommand       \teff     {{T_{\mathrm{eff}}}}
\newcommand		  \K        {\,{\rm K}\,}
\newcommand       \Rsq      {{R^2}}
\newcommand       \mum      {{\,\mu\mathrm{m}}}
\newcommand       \Rnoise   {{R^2_{\mathrm{noise}}}}

\newcommand       \FeH		{{\mathrm{[Fe/H]}}}

\definecolor{linkcolor}{rgb}{0,0,0.25}
\hypersetup{
  colorlinks=true,        
  linkcolor=linkcolor,    
  citecolor=linkcolor,    
  filecolor=linkcolor,    
  urlcolor=linkcolor      
}

\title[Dimensionality of chemical space]{The dimensionality of stellar chemical space using spectra from the Apache Point Observatory Galactic Evolution Experiment}

\author[Price-Jones \& Bovy]{
Natalie Price-Jones,$^{1,2}$\thanks{E-mail: price-jones@astro.utoronto.ca}
Jo Bovy$^{1,2,3,4}$
\\
$^{1}$Department of Astronomy and Astrophysics, University of Toronto, 50 St. George Street, Toronto ON M5S 3H4, Canada\\
$^{2}$ Dunlap Institute for Astronomy and Astrophysics, University of Toronto, 50 St. George Street, Toronto, ON M5S 3H4, Canada\\
$^{3}$ Center for Computational Astrophysics, Flatiron Institute, 162 5th Ave, New York, NY 10010, USA\\
$^{4}$ Alfred P. Sloan Fellow  
}

\date{Accepted XXX. Received YYY; in original form ZZZ}

\pubyear{2017}

\begin{document}
\label{firstpage}
\pagerange{\pageref{firstpage}--\pageref{lastpage}}
\maketitle

\begin{abstract}
	Chemical tagging of stars based on their similar compositions can offer new insights about the star formation and dynamical history of the Milky Way. We investigate the feasibility of identifying groups of stars in chemical space by forgoing the use of model derived abundances in favour of direct analysis of spectra. This facilitates the propagation of measurement uncertainties and does not presuppose knowledge of which elements are important for distinguishing stars in chemical space. We use $\sim$16,000 red-giant and red-clump $H$-band spectra from the Apache Point Observatory Galactic Evolution Experiment and perform polynomial fits to remove trends not due to abundance-ratio variations. Using expectation maximized principal component analysis, we find principal components with high signal in the wavelength regions most important for distinguishing between stars. Different subsamples of red-giant and red-clump stars are all consistent with needing about 10 principal components to accurately model the spectra above the level of the measurement uncertainties. The dimensionality of stellar chemical space that can be investigated in the $H$-band is therefore $\lesssim 10$. For APOGEE observations with typical signal-to-noise ratios of 100, the number of chemical space cells within which stars cannot be distinguished is approximately $10^{10\pm2} \times (5\pm 2)^{n-10}$ with $n$ the number of principal components. This high dimensionality and the fine-grained sampling of chemical space are a promising first step towards chemical tagging based on spectra alone.  
\end{abstract}


\section{Introduction}

In general, the observational study of galaxy evolution is statistical, using observations of many different galaxies to constrain the general behaviour of a larger population with shared characteristics (e.g., \citealt{VanDokkum2013}). In the Milky Way, we have the unique opportunity to contrast these probabilistic models with a detailed evolutionary history informed by observations of individual stars. Tracing a star through the Milky Way's evolution has become increasingly possible thanks to the many large surveys of the Milky Way (e.g., RAVE - \citealt{Steinmetz2006}; LAMOST - \citealt{Zhao2012}; APOGEE - \citealt{Majewski2015}; GALAH - \citealt{DeSilva2015}; Gaia - \citealt{GaiaCollaboration2016}). However, a star's path through the Galaxy cannot be traced through kinematic information alone, as gravitational interactions throughout its life erase its orbital history. Although stars mostly form together in chemically homogeneous clusters (\citealt{DeSilva2006}, \citealt{DeSilva2007}, \citealt{Bovy2016}), most clusters are dispersed by gravitational interactions on timescales of $< 100$ Myr \citep{Lada2003}. 

Unlike kinematic properties, the surface chemical composition of most stars evolves predictably over a stellar lifetime (e.g., \citealt{Kraft1994}, \citealt{Weiss2000}). Accurate measurements of this composition for many stars may be able to partially reconstruct information about stellar formation environments. We can take advantage of this by employing chemical tagging, the process of grouping stars based on their positions in chemical abundance space that was first proposed by \citet{Freeman2002}. Successfully chemical tagging in the weak limit---looking for large scale patterns in chemical space---requires precise spectroscopic measurements of a large sample of stars. With an appropriate sample, this form of chemical tagging can identify trends in abundances with stellar age, Galactocentric height or radius (e.g., \citealt{Haywood2013}, \citealt{Hayden2015}, \citealt{Anders2016}, \citealt{Bovy2015}, \citealt{Fernandez-Alvar2016}), and reveal chemical subpopulations in large-scale components of the Galaxy (e.g., \citealt{Martell2010}, \citealt{Schiavon2015}, \citealt{Recio-Blanco2017}) or within smaller structures such as globular clusters (e.g., \citealt{Schiavon2016}, \citealt{Tang2017}). One can also perform chemical tagging in the weak regime to investigate the chemical properties of larger structures like the Galactic disk or halo previously identified with kinematic information (e.g., \citealt{Hawkins2015}, \citealt{Wojno2016}).


The strong limit of chemical tagging is the powerful process of finding `birth clusters': identifying groups of stars in chemical space that were born in the same giant molecular cloud (GMC) without using any dynamical information. Success in this limit would allow us to address detailed questions about the Milky Way's star formation and enrichment history as well as probing stellar migration after birth and providing an additional measure of stellar ages (\citealt{Bland-Hawthorn2010}, \citealt{Mitschang2014}). Whether strong chemical tagging is possible in practice depends on the physics of star formation. The first requirement for strong chemical tagging is that the star-forming GMC is well mixed (see \citet{Feng2014} for a turbulent mixing model) and not enriched during formation such that the resulting cluster is chemically homogeneous. Recent work on open clusters and moving groups has shown that these as yet undispersed birth clusters are homogeneous to the level of our measurement precision (e.g. \citealt{DeSilva2006}, \citealt{DeSilva2007}, \citealt{Bovy2016}). It is unlikely that these clusters are truly homogeneous; recent work by \citet{Liu2016} found abundance variations between pairs of stars belonging to the same open cluster. However if these variations are sufficiently small, or confined to a small subset of cluster stars, a birth cluster will still appear as an overdensity in chemical space. The second requirement for strong chemical tagging is that the star-forming GMCs from which birth clusters form are chemically distinct, giving each birth cluster a unique chemical signature. The question of birth cluster uniqueness has been addressed by \citet{Blanco-Cuaresma2015}, who found significant overlap of some chemical abundances for their sample of open clusters. However, they were able to identify elements that allowed for a greater degree of discrimination between the clusters, implying that probing the appropriate spectral features may still provide unique chemical signatures for each cluster.

As strong chemical tagging offers a new window into the Milky Way's history, there have been many recent attempts to determine its viability. Some blind chemical tagging studies have offered promising results (e.g. \citealt{Hogg2016}, \citealt{Jofre2016}) for identifying groups of stars from chemical space information alone. However not all approaches have been optimistic about the uniqueness of cluster chemical signatures and the dimensionality of the space they span (e.g. \citealt{Mitschang2014}, \citealt{Ting2015a}, \citealt{Blanco-Cuaresma2016}, \citealt{Ness2017}). 

These studies share a similar approach to chemical tagging by making use of model-derived chemical abundances. This approach to measuring a star's chemistry is known to be problematic; most abundance determinations use simplified models that do not fully capture the complexity of stellar photospheres, and chemical abundances are derived simultaneously with other parameters with which they are degenerate. This leads to residual systematic trends between abundances and other stellar parameters (e.g., \citealt{Holtzman2015}). To avoid this, other approaches for finding accurate chemical abundances have been developed (e.g, \citealt{Ness2015}, \citealt{Rix2016}). Here, we propose to circumvent the problem of precise abundance measurements by using stellar spectra directly. This allows a more transparent approach to tracking measurement uncertainties, and does not rely on prior knowledge of which elements are relevant in the spectrum. Because many parts of a spectrum are affected by the same element, or by elements that share production channels, parts of the spectrum will be correlated and thus not all of a star's spectrum uniquely differentiates it from other stars. Given this, we examine the spectral space and reduce its dimensionality to orthogonal directions that maximally explain variance between spectra above the level of the measurement uncertainty. We determine that the resulting space is high dimensional ($n \approx10$ dimensions) and well sampled ($\approx 5$ chemical space cells per dimension), which has promising implications for the future ability to identify birth clusters in this space.

We begin our approach to this result by describing our dataset of $H$-band red giant spectra from the Apache Point Galactic Evolution Experiment (APOGEE; \citealt{Majewski2015}) in \S\ref{sec:data}. We continue in \S\ref{sec:methods} by outlining our approach to analyzing this data, describing the masking of bad pixels, the removal of bulk stellar properties through polynomial fitting, and our use of Expectation Maximized Principal Component Analysis (EMPCA) to reduce the spectra to only their most relevant principal components. In \S\ref{sec:app}, we show the results of applying our technique to open cluster, red clump, and red giant stars, finding the number of relevant components and thus the dimensionality and granularity of the chemical space spanned by each sample. We describe the consequences of these results and avenues for future work in \S\ref{sec:discussion}, concluding with a brief summary in \S\ref{sec:conclusion}.

\begin{figure*}
\centering
\includegraphics[width = \linewidth]{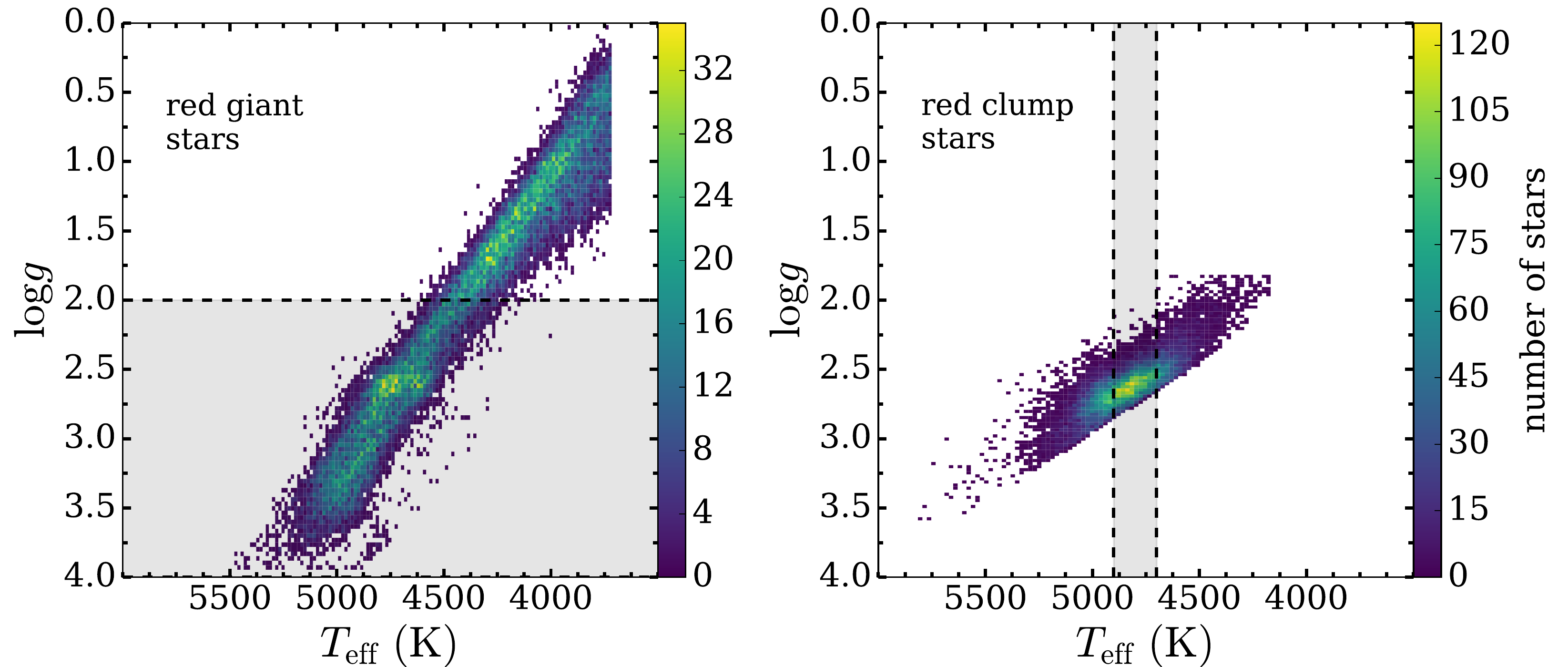}
\caption{Spectroscopic Hertzsprung-Russell diagram of the two main samples used in this paper. The left panel shows the red giant stars, while the right panel shows the red clump stars. Both distributions peak around $\log g=2.5$ and $\teff=4750\K$.The red giant sample contains 26,923 stars, while the red clump sample contains 19,937 stars. Shaded regions show the stars used in this analysis.}
\label{fig:HR}
\end{figure*}

\section{Data Set}
\label{sec:data}
The high resolution ($R\sim 22,500$) stellar spectra used in this study were taken by APOGEE, an instrument mounted on the Sloan Digital Sky Survey (SDSS - \citealt{Eisenstein2011}). APOGEE is an H-band ($1.5\mum$ to $1.7\mum$) spectrograph capable of simultaneously observing 300 targets \citep{Majewski2015}. The survey's target selection populates fields that span up to $\sim$5 kpc away from the Earth in the Milky Way's disk and up to $\sim$1 kpc away from the Galactic midplane \citep{Zasowski2013}. This multi-object spectroscopy is combined with the 3 degree field of view of the Apache Point Observatory 2.5 m telescope \citep{Gunn2006} in order to obtain broad spatial coverage of the Milky Way and to allow for repeated observations of most of the $150,000$ stars in the APOGEE sample. 

These repeated observations are radial velocity corrected, roughly continuum normalized, then combined into a single spectrum by the data reduction pipeline (\citealt{Nidever2015}). The APOGEE Stellar Parameter and Chemical Abundances Pipeline (ASPCAP; \citealt{Perez2015}) continuum-normalizes the combined spectra and fits with a grid of synthetic spectra \citep{Zamora2015} to determine the stars' effective temperature $\teff$, surface gravity $\log g$, and metallicity $Z$, as well as the abundances of 15 elements with absorption features in the H-band (C, N, O, Na, Mg, Al, Si/H, S/H, K/H, Ca/H, Ti/H, V/H, Mn/H, Fe, Ni - \citealt{Holtzman2015}). We use data from APOGEE's public data release 12 (DR12 - \citealt{Alam2015}) to avoid the artificial adjustments to the measurement uncertainties used to strongly downweight persistence regions that are present in subsequent data releases.

Of the 150,000 stars in the APOGEE survey, we use two subsamples: red-giant stars and red-clump stars. Each sample consists of $\sim 20,000$ stars, plotted in $\log g$ and $\teff$ in Figure~\ref{fig:HR}. Stars belonging to each sample were selected by a cut in the stellar properties provided by the ASPCAP. The selection process for the red-clump in APOGEE DR12 combines ASPCAP stellar parameters with the results of simulated stellar evolution, and is described in detail in \citet{Bovy2014}. To select the red-giant subsample, we first constrain metallicity $Z > -0.8$. Of this subsample we then choose the objects that are either bluer than the red clump ($(J - K_s)_0 \geq 0.8$) or that do not satisfy the red clump constraint derived in \citet{Bovy2014}:
\begin{eqnarray}
	\log g &>& 0.0018\, \mathrm{dex}/\mathrm{K} \left(T_{\mathrm{eff}} - T_{\mathrm{eff}}^{\mathrm{ref}}([\mathrm{Fe/H}])\right)+2.5	\nonumber\\
	T_{\mathrm{eff}}^{\mathrm{ref}} &=& -382.5\,\mathrm{K}/\mathrm{dex}\, [\mathrm{Fe/H}] + 4607 \, \mathrm{K}
\end{eqnarray}
$J$ and $K_s$ are colours from 2MASS \citep{Skrutskie2006}, and the zero subscript indicates that colours are dereddened. By applying these selection cuts to APOGEE targets with asteroseismology in the APOKASC catalog \citep{Pinsonneault2014} for which the evolutionary state is known from the observed frequency and period spacing of stellar oscillations, we find that these red-giant selection cuts create a red-giant sample with only $\sim 3\%$ contamination from red-clump stars.

The choice of these samples was advantageous for our goal of finding the dimensionality of spectral space for several reasons. Red giants are intrinsically luminous, which allows them to be observed at high signal-to-noise ratio over a wide area of the Galaxy, and they have cool photospheres, which causes many absorption features to be present in their spectra. They are also known to rotate slowly (\citealt{Gray1982}, \citealt{DeMedeirosJ.R.1996}) and the effect of stellar rotation on spectral features is therefore minimal. An additional advantage of using red giant stars was highlighted in the work of \citet{Dotter2017} on surface abundance evolution from initial bulk abundances through atomic diffusion. \citet{Dotter2017} showed that changes between initial and surface abundances are small for red giants, and argued that comparing surface abundances for stars in the same evolutionary phase should reduce the uncertainty introduced by the time evolution of surface abundances. 

In addition to the two main data samples, red clump and red giant stars belonging to a set of open clusters are used as a test case of the methods outlined in the following section. Open clusters that were sufficiently sampled by APOGEE ($\geq$ 10 stars) were chosen for this purpose: NGC 6819, NGC 2158 and M67. We also tested our methods on globular cluster M13, which is known to have multiple populations in chemical space \citep{Carretta2009}. We obtain members of these clusters from the work of \citet{Meszaros2013} and \citet{Meszaros2015}.


\section{Methods}
\label{sec:methods}

Abundances are an obvious way to access chemical space. However, standard methods for calculating abundances involve fitting complex model spectra with a large number of parameters. Varying these parameters to optimize a fit, even assuming a one-dimensional star and local themodynamic equilibrium, is computationally taxing and simplifying assumptions are made (\citealt{Smiljanic2014}, \citealt{Perez2015}). The complexity of the fitting procedure and the assumptions on the photosphere needed to make a fit possible make deriving realistic measurement uncertainties very challenging, which in turn makes it difficult to assess the precision of the resulting abundances. This challenge has inspired new approaches to measuring chemical abundances, including polynomial fitting of stellar properties to determine abundances \citep{Rix2016}, linear interpolation of abundances from an existing non-rectangular grid of model spectra \citep{Ting2016}, and the use of machine learning algorithms to predict abundances based on training data where abundances are well known (\citealt{Ness2015}, \citealt{Casey2016a}). While these methods offer improved precision in chemical abundances, they still aim to calculate abundances for a fixed number of elements. For our work assessing the dimensionality of chemical abundance space, we are only interested in the variation between stellar spectra due to their differing chemistry, not the chemical abundances themselves. Because of this, we are able to forgo computing these abundances in favour of a simpler method that reveals chemical information despite relying on a few assumptions: (a) that non-chemical differences between stars are fully described by specifying $\teff$, $\log g$ and [Fe/H], (b) that the overall dependence of the spectra on $\teff$, $\log g$ and [Fe/H] can be described using quadratic functions of these quantities, (c) that the spectral differences coming from varying the abundance ratios at fixed ($\teff$, $\log g$, [Fe/H]) are essentially linear, with no dependence on those quantities, and finally (d) that the noise model associated with spectra from the APOGEE survey serves as an accurate description of the measurement uncertainties. As we demonstrate with our results, these assumptions approximately hold for the relatively narrow temperature range of the samples we consider. We outline below our approach to isolating the differences between stellar spectra due only to their differing chemical abundances, a method first described in \citet{Bovy2016}. 

In this section, we use the following notation conventions. Matrices are bolded upper case characters (e.g., $\mathbf{V}$), vectors are bolded lower case characters (e.g., $\mathbf{v}$) and scalars are lower case characters (e.g., $v$). Columns or rows of a matrix are denoted as bold lower case characters with the index of their row or column: $\mathbf{v}^{i}$ is the $i$'th row of $\mathbf{V}$, and $\mathbf{v}_j$ is the $j$'th column of $\mathbf{V}$. Individual elements of a matrix are identified by their indices; the jth element of the ith row of a matrix $\mathbf{V}$ is denoted $\mathbf{V}_{ij}$. The symbol $\cdot$ denotes matrix multiplication, while the symbol $*$ indicates element-wise multiplication. Finally, the index $s$ is used for dimensions that span the number of stars in the sample, from $1$ to $S$, the index $p$ is used for dimensions that span the number of pixels in a spectrum, from 1 to $P$, and the index $n$ is used for dimensions that span the number of principal components used to decompose the sample.

The code used to implement the methods in the following sections is available online through the \texttt{spectralspace} Python package, which can be installed from \url{https://github.com/npricejones/spectralspace}.

\subsection{Pixel level masking of spectra}

We begin by masking untrusted parts of each individual spectrum using the APOGEE\_PIXMASK bitmask from the ASPCAP \citep{Holtzman2015}. In our `standard mask', we exclude pixels where bits 0, 1, 2, 3, 4, 5, 6, 7 or 12 are set. In our more aggressive `persistence mask', we additionally exclude pixels where bits  9, 10, or 11 are set. The full breakdown of the APOGEE\_PIXMASK bitmask can be found in Table~\ref{tab:bitmask}.


\begin{table}
	\caption{APOGEE\_PIXMASK bitmask flags and their use in our masks \citep{Holtzman2015}.}
	\label{tab:bitmask}
	\begin{tabular}{|c|}
		\textbf{Standard mask}\\
	\end{tabular}
	\\
	\begin{tabular}{|c|l|} 
		\hline
        Bit & Flag  \\
        \hline
        0 & Pixel marked as bad according to bad pixel mask \phantom{here is some ti}\\
        1 & Pixel struck by cosmic ray \\
        2 & Pixel saturated \\
        3 & Pixel marked as unfixable\\
        4 & Pixel marked as bad according to dark frame \\
        5 & Pixel marked as bad according to flat frame \\
        6 & Pixel set to have a very high error\\
        7 & No sky available for this pixel from sky fibers\\
        12 & Pixel falls near sky line \\
        \hline
	\end{tabular}
	\\
	\begin{tabular}{|c|}
		\textbf{Persistence mask (Standard mask with the following additional bits)}\\
	\end{tabular}
	\\
	\begin{tabular}{|c|l|} 
		\hline
        Bit & Flag  \\
        \hline
        9 & Pixel falls in high persistence region \phantom{here is some texty stuff nee}\\
        10 & Pixel falls in medium persistence region \\
        11 & Pixel falls in low persistence region \\
        \hline
    \end{tabular}
    \\
	\begin{tabular}{|c|}
		\textbf{Unused bits}\\
	\end{tabular}
	\\
	\begin{tabular}{|c|l|} 
		\hline
        Bit & Flag  \\
        \hline
        8 & Pixel falls in Littrow ghost \\
        13 & Pixel falls near telluric line \\
        14 & Less than half of star's PSF is seen in good pixels \phantom{here is some t}\\
        \hline
	\end{tabular}
\end{table}

In most samples, we also apply a cut to remove stars that were frequently observed with fibers known to be affected by super-persistence in the blue detector, which mostly affects fibers 0-100. After applying one of the masks and potentially applying a fiber cut, we also mask on individual pixels with low signal-to-noise ratio, calculated as SNR = $\mathbf{F}/\mathbf{M}$, where $\mathbf{F}$ are the continuum-normalized spectra and $\mathbf{M}$ are the measurement uncertainties. We mask pixels with SNR $< 50$. After this masking, we check that each pixel $p$ has enough unmasked stars to do a second order polynomial fit (i.e., more than fifteen stars for the red clump and red giant subsamples, or more than five stars for the clusters). Pixels that do not meet this requirement are masked for all stars in the sample, although this condition only triggered in the smaller open cluster samples.

\subsection{Removal of non-abundance ratio spectral variations}

\begin{figure}
\centering
\includegraphics[width = \linewidth]{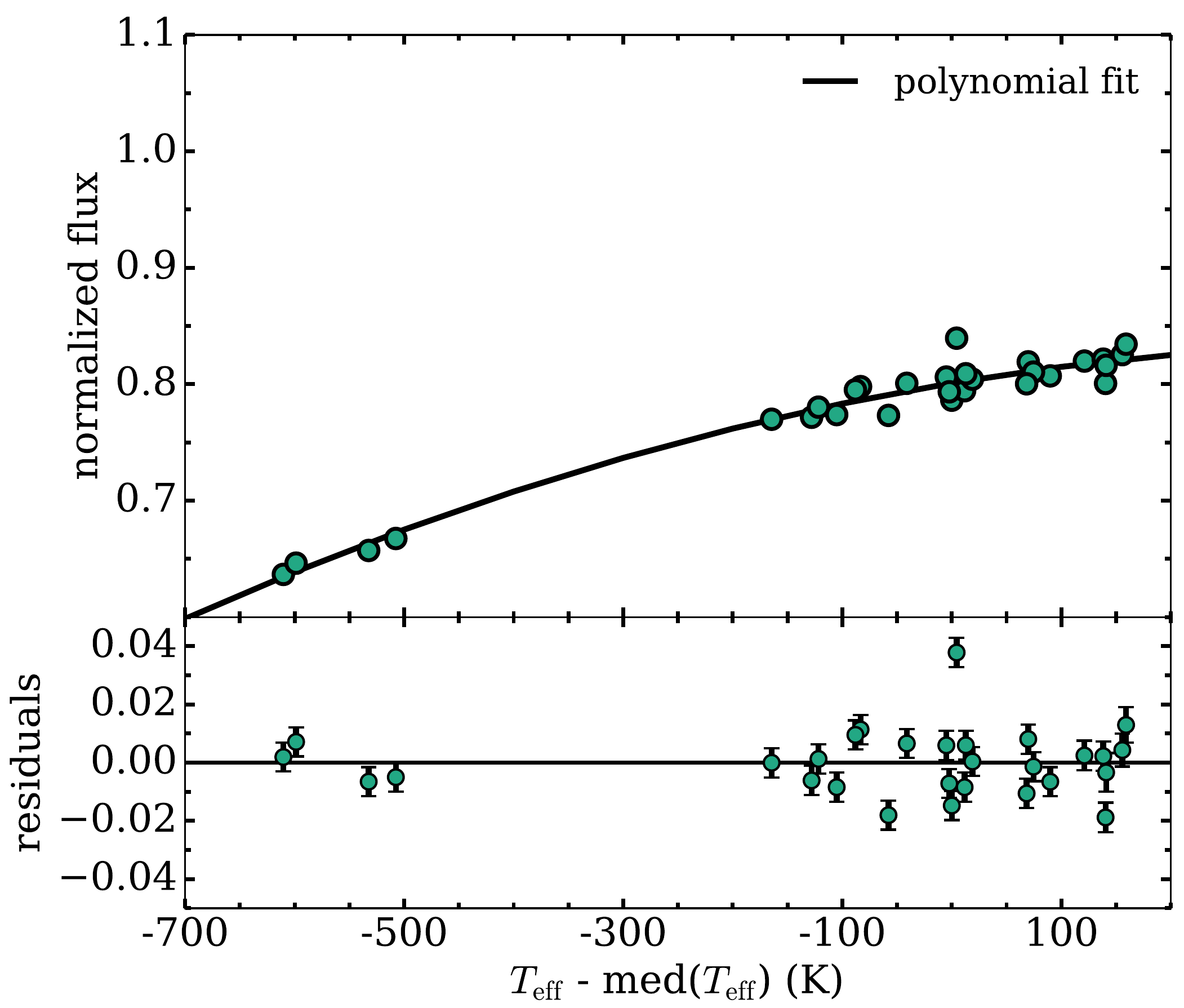}
\caption{ An example of a fit to remove the overall temperature trend for the stars of open cluster NGC6819 at a pixel corresponding to a carbon absorption feature (pixel 4313, $1.618997\mum$). The x-axis is effective temperature normalized to the median effective temperature (4731 K), while the y-axis in both plots is continuum-normalized flux. The top panel shows the result of the fit, and the bottom shows the residuals from the fit.}
\label{fig:fit}
\end{figure}

To remove overall non-chemical trends with temperature, gravity and metallicity from the stellar spectra, we compute a polynomial fit at each of the $P$ pixels in the spectra. Specifically, we fit in effective temperature $\teff$, surface gravity $\log g$ and iron abundance [Fe/H], where these quantities are as computed by the ASPCAP. This choice of parameters captures the primary properties of a given star; along the giant branch they are convenient proxies for mass, age and metallicity. For a given pixel $p$, we model the flux at that pixel as a second order polynomial. We compute the best-fitting coefficients $\mathbf{b}_p$ by solving the matrix equation $\mathbf{f}_{p} = \mathbf{X}\mathbf{b}_{p}$, where $\mathbf{f}_p$ is the $p$'th column vector of $\mathbf{F}$, $\mathbf{b}_{p}$ is a column vector corresponding to the fit coefficients and $\mathbf{X}$ is a two dimensional matrix of fit variables constructed as the transpose of Equation~\eqref{eqn:Xmatrix}:
\begin{equation}
\begin{aligned}
\begin{split}
\mathbf{X}^{\mathrm{T}} & =
\begin{pmatrix}
{\teff_{,0}}^2 & \cdots & {\teff_{,\mathrm{S}}}^2\\
\teff_{,0} & \cdots & \teff_{,\mathrm{S}}\\
(\log g)_0^2 & \cdots & (\log g)_{\mathrm{S}}^2\\
(\log g)_0 & \cdots & (\log g)_{\mathrm{S}}\\
\FeH_0^2 & \cdots & \FeH_{\mathrm{S}}^2\\
\FeH_0 & \cdots & \FeH_{\mathrm{S}}\\
\teff_{,0}*(\log g)_0 & \cdots & \teff_{,\mathrm{S}}*(\log g)_{\mathrm{S}}\\
\teff_{,0}*\FeH_0 & \cdots & \teff_{,\mathrm{S}}*\FeH_{\mathrm{S}}\\
(\log g)_0*\FeH_0 & \cdots & (\log g)_{\mathrm{S}}*\FeH_{\mathrm{S}}\\
1 & \cdots & 1
\end{pmatrix}
\end{split}.
\label{eqn:Xmatrix}
\end{aligned} 
\end{equation}

\begin{figure*}
\centering
\includegraphics[width = \linewidth]{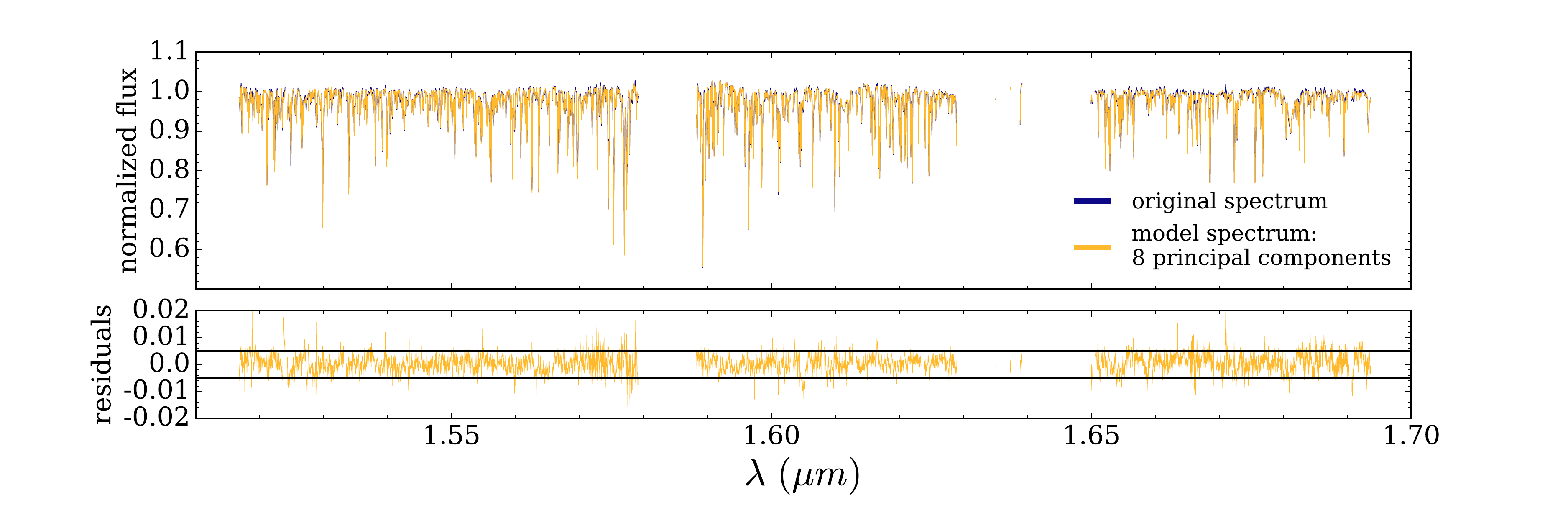}
\caption{Modelling a spectrum using the first eight principal components identified by EMPCA. The top panel shows the original spectrum with a reconstructed model overplotted. The bottom panel shows the residuals between the original spectrum and model with horizontal lines marking the median uncertainty for this star (0.005). The gaps from $\sim$1.58 to $\sim$1.59 $\mu$m and $\sim$1.64 to $\sim$1.65 $\mu$m are the spaces between the APOGEE detectors. The gap from $\sim$1.63 to $\sim$1.64 $\mu$m is an area of the `green detector' seriously effected by persistence and masked in our analysis.}
\label{fig:model}
\end{figure*}

\begin{figure}
\centering
\includegraphics[width = \linewidth]{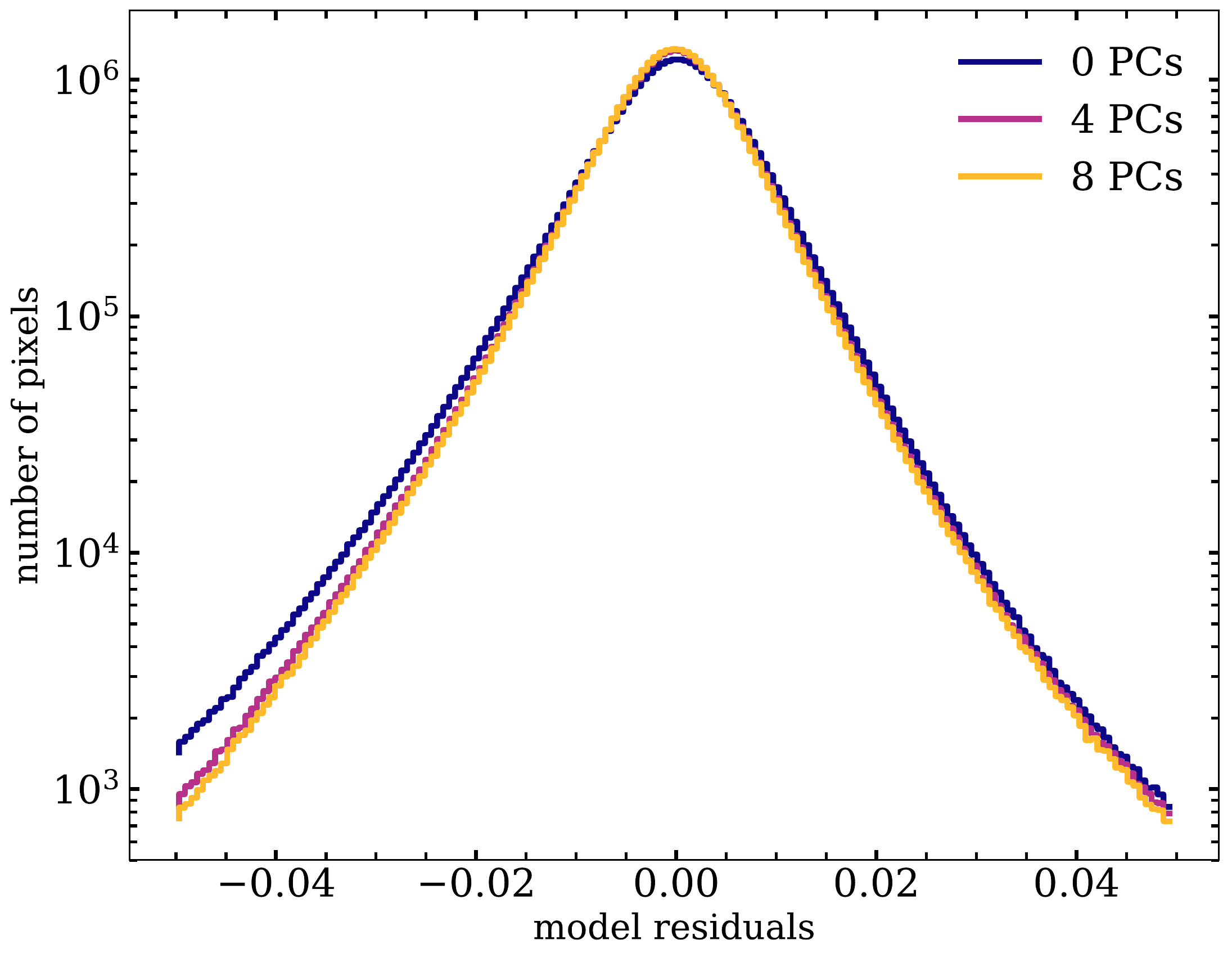}
\caption{Histograms of the residual normalized flux after subtracting from each spectrum in the example slice a model using 0, 4 and 8 PCs. As more PCs are used, the histogram becomes more sharply peaked in the centre, with shrinking wings.}
\label{fig:residualhist}
\end{figure}

This fit is overconstrained and the measurements $\mathbf{y}_{p}$ have associated uncertainties. We define the best fit as those coefficients that minimize 
\begin{equation}
	\chi^2 = \left(\mathbf{f}_p - \mathbf{X}\mathbf{b}_p\right)^T \mathbf{V}_p^{-1} \left(\mathbf{f}_p - \mathbf{X}\mathbf{b}_p\right)
\end{equation}
where $\mathbf{V}_p$ is the covariance matrix of the measurement uncertainties. The minimum $\chi^2$ occurs when
\begin{equation}
\mathbf{b}_{p} = \left[\mathbf{X}^{\mathrm{T}} \cdot \mathbf{V}_{p}^{-1} \cdot \mathbf{X}\right]^{-1} \cdot \left[\mathbf{X}^{\mathrm{T}} \cdot \mathbf{V}_{p}^{-1} \cdot \mathbf{f}_{p}\right],
\label{eqn:coeffs}
\end{equation}
In the simplest case (which we assume here), $\mathbf{V}_p$ is diagonal with diagonal elements $\mathbf{m}_p * \mathbf{m}_p$ where $\mathbf{m}_p$ is, the $p$'th column of the measurement uncertainties $\mathbf{M}$. If the simplest case cannot be assumed, $\mathbf{V}_p$ must remain the full covariance matrix. While we ignore correlations between pixels for the purposes of this study, these correlations will impact the principal components we derive in \S\ref{sec:dimension}, making the number we identify as important for modelling spectra an upper limit.

With the coefficients $\mathbf{b}_{p}$, we can find the residuals for pixel $p$ 
\begin{equation}
	\mathbf{d}^p = \left[\mathbf{f}_{p} - \mathbf{X}\mathbf{b}_{p}\right]^{\mathrm{T}}	
\end{equation}
where $\mathbf{d}^p$ is the $p$'th row of the residual matrix $\mathbf{D}$. The matrix $\mathbf{D}$ with rows $\mathbf{d}^p$ is a new version of the spectra in our sample in which the effects of $\teff$, $\log g$ and [Fe/H] (representing mass, age and metallicity) have been removed.

An example of the results of the fitting process on stars in open cluster NGC6819 is shown in Figure~\ref{fig:fit}. Given that the range of age and metallicity within a cluster are small, we use only $\teff$ in the polynomial fit. The top panel shows the fit compared to the data used to generate it, while the lower panel displays the resulting residuals at that pixel.

\subsection{Dimension reduction}
\label{sec:dimension}
Our goal now is to analyze the fit residuals from all pixels to understand how they inform us about the dimensionality of the chemical space spanned by the sample. Because we have removed the overall trends with mass, age and metallicity from the spectra, we work under the assumption that any remaining variation between the spectra is due to their having different abundance ratios $[\mathrm{X/Fe}]$. We discuss the caveats to and limitations of this assumption in \S\ref{sec:discussion} below.

Since not every one of the $P$ pixels is relevant for distinguishing between stars, we employ an algorithm that will reduce this number of dimensions to only those that are relevant. Classical principal component analysis (PCA; \citealt{Joliffe2002}, \citealt{Ivezic2014}) solves exactly this problem by minimizing
\begin{equation}
\chi^2 = \sum_{p=1,s=1}^{P,S}\left[\mathbf{D}_{ps} - \sum_{n=1}^{N}\left(\mathbf{E}_{pn}*\mathbf{C}_{ns}\right)\right]^2,
\end{equation}
where N is the number of spectra, $\mathbf{D}$ are the fit residual spectra, $\mathbf{E}$ is a matrix whose columns are the eigenvectors (or principal components) of $\mathbf{D}$'s covariance matrix, $N$ is the total number of components in the model, and $\mathbf{C}$ is the matrix of coefficients that scale the principal components to model $\mathbf{D}$.


In our case, the resulting principal components will have the greatest magnitude in pixels that are most important for distinguishing between stellar spectra. PCA assumes the spectra vary linearly, and so a linear combination of principal components with appropriate coefficients should be an accurate model of the input spectra. This approximately applies to our data set of red giants in a fixed temperature range, where intensity in a pixel across all stars in the sample is well modelled by a second order polynomial (see Figure~\ref{fig:fit} for an example fit). Subtracting this polynomial leaves small residuals, which can be well approximated as linear variation (see Figure~\ref{fig:model} for an example of a spectrum reconstructed from principal components and the polynomial fits for each pixel). 

\begin{figure*}
\centering
\includegraphics[width = \linewidth]{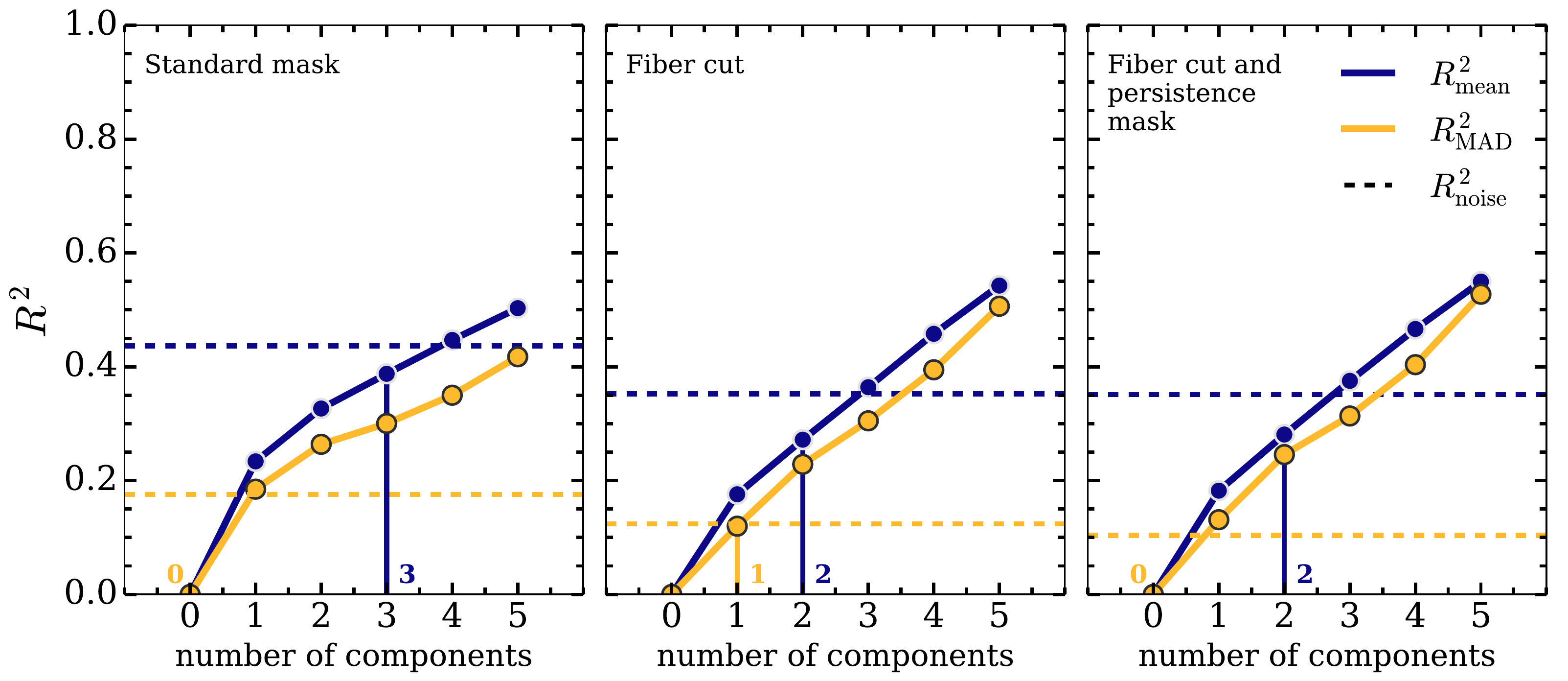}
\caption{Each panel shows an $\Rsq$ curve for open cluster NGC 6819 using two methods of computing the statistic (variance and median absolute deviation). Dashed horizontal lines indicate the corresponding value of $\Rnoise$. The left most panel is the result of computing $\Rsq$ using the standard bitmask (see Table~\ref{tab:bitmask} for a summary of bits set in this mask). The central panel is the result of cutting out stars observed with APOGEE fibers known to be effected by persistence (the 'blue' part of the detector). The rightmost panel is the result of cutting out the bad fibers and applying a more aggressive mask to remove persistence-effected regions outside of the blue part of the detector.}
\label{fig:n6819R2}
\end{figure*}

However, classical PCA has some limitations, in that it weights all data equally and has no way to handle missing data. Our dataset has already been masked to remove untrusted data, and has measurement uncertainties that should be incorporated to weight against finding directions of data variance that are due to noise. To retain the dimension reducing power of PCA but avoid these limitations, we employed expectation maximized PCA (EMPCA; \citealt{Dempster1977}, \citealt{Roweis1997}). We use the implementation from \citet{Bailey2012} and refer the reader to that paper for a detailed description of the algorithm summarized below.

\subsubsection{EMPCA algorithm}
\label{sec:EMPCA}

EMPCA seeks to find the set of principal components that maximizes the likelihood of that model being an accurate description of the data given uncertainties in that data.  The algorithm minimizes
\begin{equation}
	\chi^2 = \sum_s^S\left[\mathbf{d}_s - \mathbf{E}\cdot\mathbf{c}_{s}\right]^T\cdot\mathbf{V}_{s}^{-1}\cdot\left[\mathbf{d}_s - \mathbf{E}\cdot\mathbf{c}_{s}\right]
	\label{eqn:chisquared}
\end{equation}
where $\mathbf{d}_s$ is the $s$'th column of residual matrix $\mathbf{D}$, $\mathbf{c}_s$ is the $s$'th column of the coefficient matrix $\mathbf{C}$ and $\mathbf{V}_{s}$ is the pixel covariance matrix for star $s$. The algorithm minimizes this objective function by starting with a random set of principal components for the columns of $\mathbf{E}$, then iterating over the following two steps:
%
\begin{enumerate}
\item The expectation step: Fixing $\mathbf{E}$, solve for the coefficient matrix $\mathbf{C}$. We find the columns of $\mathbf{C}$ that minimize Equation~\eqref{eqn:chisquared} with
\begin{equation}
	\mathbf{c}_s = \left[\mathbf{E}^{\mathrm{T}}\cdot\mathbf{V}_{s}^{-1}\cdot \mathbf{E}\right]^{-1}\cdot\left[\mathbf{E}^{\mathrm{T}}\cdot\mathbf{V}_{s}^{-1}\cdot\mathbf{d}_s\right]
\end{equation}
%

\item The maximization step: Use the solution for $\mathbf{C}$ to update $\mathbf{E}$. We solve for each column of $\mathbf{E}$ after subtracting out a model constructed from the previous columns. Thus each column of $\mathbf{E}$ is given by
\begin{equation}
	\mathbf{e}_n = \left[\sum_s^S \mathbf{C}_{ns} * \mathbf{V}_{s}^{-1} * \mathbf{C}_{ns}\right]^{-1}\cdot\left[\sum_s^S \mathbf{C}_{ns} * \mathbf{V}_{s}^{-1} \cdot \mathbf{d}^{(n)}_s\right]
\end{equation}
%
where $\mathbf{d}^{n}_s$ is the $s$'th column of $\mathbf{D}^{(n)}$, whose elements are defined as
\begin{equation}
\mathbf{D}^{(n)}_{ps} = \mathbf{D}_{ps} - \sum_{i<n}\mathbf{e}_{pi}\cdot\mathbf{c}_{is},
\label{eqn:deltaj}
\end{equation}
i.e. the fit residual spectra after the first $n-1$ principal components have been subtracted. 
\end{enumerate}

The quality of the resulting model is demonstrated in Figure~\ref{fig:model}, where an input red-clump stellar spectrum is reconstructed by multiplying the first 8 principal components by their  coefficients for the chosen star. After summing these coefficient-scaled principal components, we add back the values of polynomial fits at each pixel and plot the resulting model over the original spectrum. In the lower panel of Figure~\ref{fig:model}, we show the residuals between the model and the input spectrum, with horizontal lines marking the median uncertainty about zero for this spectrum. With just 8 principal components we can model this particular spectrum to within the typical measurement uncertainty. 

In Figure~\ref{fig:residualhist}, we show the distribution of model residuals for all spectra in our example slice of the red clump for models using 0, 4 and 8 principal components. As we expect, increasing the number of principal components used in the model causes the residual histogram to become more peaked towards a value of 0, with shallower wings. Given this continual improvement in modelling, it is necessary to define a cutoff to determine when improvements to the model explain variations due to noise, not intrinsic stellar differences.

\subsubsection{Assessing principal components}

We now define some useful statistics that can be computed from the results of EMPCA, which allow us to assess the results of the algorithm. 

We define $V_{\mathrm{data}}^{\mathrm{mean}}$ as the mean based variance of $\mathbf{D}$,
\begin{eqnarray}
	\label{eqn:meanvar}
	V_{\mathrm{data}}^{\mathrm{mean}} &=& \frac{1}{P*S}\sum_{p=1,s=1}^{P,S}\left(\mathbf{D}_{ps} - \overline{\mathbf{D}}\right)^2,\\
	\overline{\mathbf{D}} &=& \frac{1}{P* S} \sum_{p=1,s=1}^{P,S} \mathbf{D}_{ps},
\end{eqnarray}
We define a similar quantity $V_{\mathrm{data}}^{\mathrm{MAD}}$ as the median absolute deviation (MAD) based variance of $\mathbf{D}$
\begin{equation}
V_{\mathrm{data}}^{\mathrm{MAD}} = 1.4826^2\sum_{p=1}^{P}\mathrm{med}\left(\left[\mathbf{d}^p - \mathrm{med}\left(\mathbf{d}^p\right)\right]^2\right),
\label{eqn:madvar}
\end{equation}
where $\mathbf{d}^p$ is the $p$'th row of $\mathbf{D}$, the operation `$\mathrm{med}$' is taking the median of the data, and the factor of $1.4826$ scales the median absolute deviation to the standard deviation for normally distributed data, i.e. $V_{\mathrm{data}}^{\mathrm{mean}} = V_{\mathrm{data}}^{\mathrm{MAD}}$ for data drawn from a Gaussian distribution.

Equations~\eqref{eqn:meanvar} and~\eqref{eqn:madvar} represent the total variance in the data. We wish to compare these to the variance after removing the principal component model produced by EMPCA. To do this, we define two additional quantities that compute the variance in the data after the contribution from the first $n$ principal components has been subtracted out:
\begin{equation}
V_{\mathrm{model}}^{\mathrm{mean}}(n) = \frac{1}{P*S}\sum_{p=1,s=1}^{P,S}\left(\mathbf{D}^{(n)}_{ps} - \overline{\mathbf{D}^{(n)}}\right)^2,
\label{eqn:meanveig}
\end{equation}
\begin{equation}
V_{\mathrm{model}}^{\mathrm{MAD}}(n) = 1.4826^2\sum_{p=1}^{P}\mathrm{med}\left(\left[\mathbf{d}^{p\,(n)} - \mathrm{med}\left(\mathbf{d}^{p\,(n)}\right)\right]^2\right).
\label{eqn:madveig}
\end{equation}
Here $\mathbf{D}^{(n)}$ is the original data set after the model using the first $n$ principal components has been removed (Equation \eqref{eqn:deltaj}), and $\mathbf{d}^{p\,(n)}$ is the $p$'th row of $\mathbf{D}^{(n)}$. Note that in the case where $n=0$, $V_{\mathrm{model}}^{\mathrm{mean}}(0) = V_{\mathrm{data}}^{\mathrm{mean}}$ and $V_{\mathrm{model}}^{\mathrm{MAD}}(0) = V_{\mathrm{data}}^{\mathrm{MAD}}$. Also note that when the number of components is maximized ($n = \min(N,P) = n_{\mathrm{limit}}$), $V_{\mathrm{model}}^{\mathrm{mean}}(n_{\mathrm{limit}}) = 0$ and $V_{\mathrm{model}}^{\mathrm{MAD}}(n_{\mathrm{limit}}) = 0$, because in that case the model can perfectly represent all of the data. In what follows, we use $V_{\mathrm{data}}$ and $V_{\mathrm{model}}$ without superscripts when either the mean-based or MAD-based versions could be used.

We compare these variances using $R^2(n)$, defined as:
\begin{equation}
\Rsq(n) = 1 - \frac{V_{\mathrm{model}}(n)}{V_{\mathrm{data}}}.
\label{eqn:Rsq}
\end{equation}
The $R^2(n)$ curve should increase in height with the number of iterations on the model components. The maximal value for $R^2(n)$ is 1, which indicates the data can be perfectly reproduced by a linear combination of $n$ components and occurs at $n = \min(N,P) = n_{\mathrm{limit}}$.

Example $R^2(n)$ curves for $n$ from 1 to 5 are shown in Figure~\ref{fig:n6819R2}, where the upper curve was computed with mean based variance and the lower curve using median absolute deviation. In this figure, the $\Rsq$ values are compared to $\Rnoise$, a statistical quantity that describes the variance in the data due to measurement noise:
\begin{equation}
\label{eqn:Rnoise}
\Rnoise = 1 - \frac{V_{\mathrm{noise}}}{V_{\mathrm{data}}},
\end{equation}
where with our assumption that pixels are independent:
\begin{equation}
V_{\mathrm{noise}} = \sum_{p=1,s=1}^{P,S} \left(\mathbf{M}_{ps} * \mathbf{M}_{ps}\right),
\end{equation}
where $\mathbf{M}$ is the matrix of measurement uncertainties. $\Rnoise$ can vary depending on our chosen form for computing $V_{\mathrm{data}}$, using either Equation~\eqref{eqn:meanvar} or~\eqref{eqn:madvar} for mean or MAD based variance respectively. The results for both methods of computation are shown as dashed lines in Figure~\ref{fig:n6819R2}, where they are colour coded to correspond to their respective solid $R^2$ curves.

\begin{figure*}
\centering
\includegraphics[width = \linewidth]{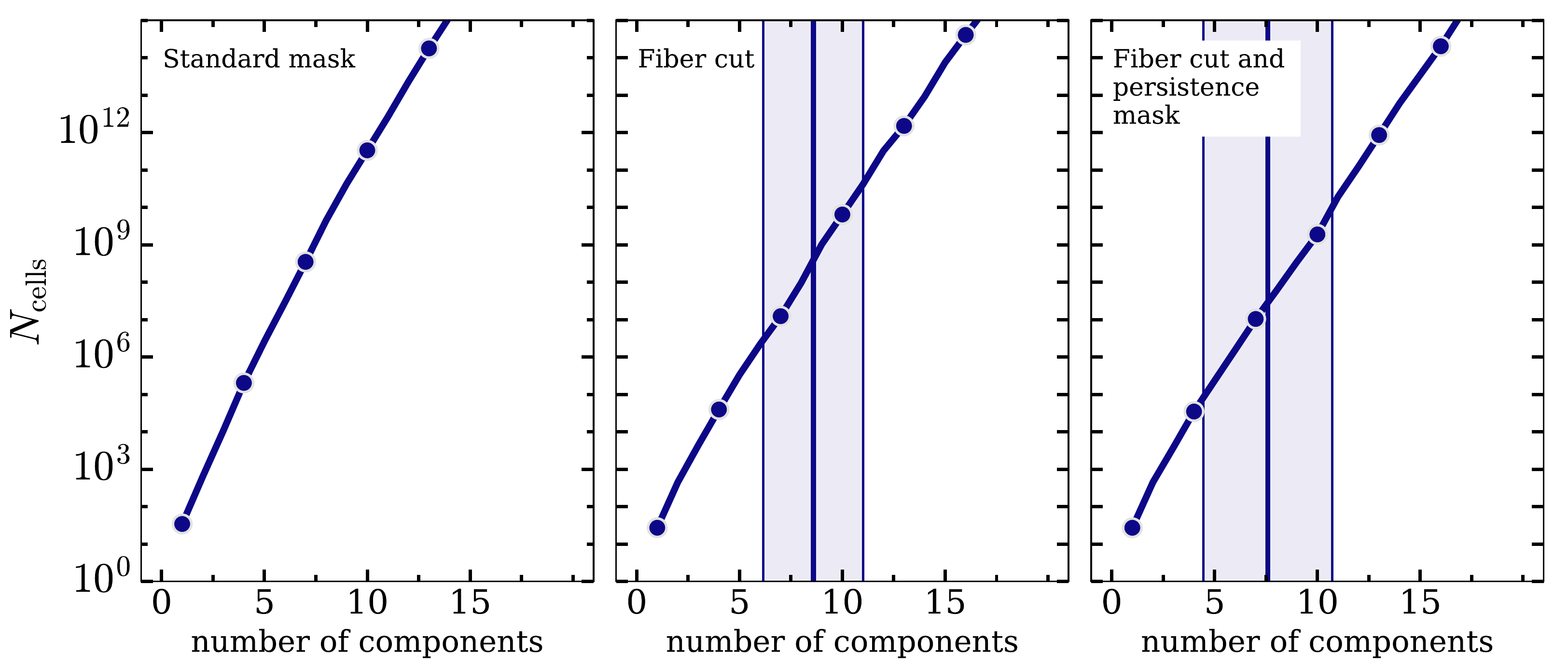}
\caption{Number of cells for the example red clump slice as a function of the number of principal components used to explain the data, computed with Equation~\eqref{eqn:Ncells} assuming Equation~\eqref{eqn:consth} for $h_n$. Vertical lines mark the average number of principal components needed to explain the data, while the shaded region is the range of 68\% confidence (see \S\ref{sec:jackknife} for how this confidence range is computed). These subplots compare different masking techniques.}
\label{fig:rcNcell}
\end{figure*}

Computing $\Rnoise$ is useful because it provides the threshold mentioned at the end of \S\ref{sec:dimension}. Adding additional principal components to a model where $\Rsq > \Rnoise$ succeeds only in explaining variation due to measurement uncertainty. We denote the maximum number of components for which $\Rsq$ is below the $\Rnoise$ threshold as $n_{\max}$.


\subsubsection{Sampling precision of chemical space}

While comparing $R^2(n)$  to $\Rnoise$ gives us a sense of how many dimensions are needed to span chemical space, it is also valuable to compute the number of chemical cells ($N_{\mathrm{cells}}$) that our analysis samples. In this we follow \citet{Ting2015a}, dividing the total chemical space volume spanned by the principal components by the volume of a single chemical space cell, which is limited by the measurement noise.

\begin{equation}
N_{\mathrm{cells}} = \frac{\mathrm{total\,\,chemical\,\,volume}}{\mathrm{chemical\,\,cell\,\,size}}.
\end{equation}

The total chemical volume is determined by the variance spanned by each principal component. In PCA, the eigenvalue  associated with $n$th principal component represents the additional variance of the data explained by adding the $n$th principal component to a model consisting of $n-1$ principal components:

\begin{equation}
\lambda_n^2 = V_{\mathrm{model}}(n-1) - V_{\mathrm{model}}(n),
\end{equation}
where $V_{\mathrm{model}}$ can be calculated with Equation~\eqref{eqn:meanveig} or Equation~\eqref{eqn:madveig}. Thus for $n$ components, the total chemical space volume corresponds to 
\begin{equation}
	\prod_{i=0}^{n} \min (S,P) \lambda_i,
\end{equation}
where $\lambda_i^2$ is the eigenvalue of the $i$th principal component, and with $7214$ pixels per APOGEE spectrum, $\min (S,P) = S$ for our samples.

The size of an individual chemical cell $h$ can be determined by how the measurement uncertainty projects onto each principal component:

\begin{equation}
h_n = \sqrt{\frac{1}{S}\sum_{s=1}^{S}\left(\mathbf{m}_{s}*\mathbf{m}_{s}\right)\cdot \mathbf{e}_n},
\end{equation}
where $\mathbf{m}_{s}$ is the $s$'th column of the measurement uncertainties $\mathbf{M}$, and $\mathbf{e}_n$ is the $n$'th principal component, the $n$'th column of $\mathbf{E}$ as defined in \S\ref{sec:methods}.
We can make more conservative estimates for $h_n$ by assuming that the noise is the essentially the same in all pixels, in which case $h_n \rightarrow h$ with $h$ given by
\begin{equation}
h = \sqrt{\frac{1}{P*S}\sum_{p=1,s=1}^{P,S} \mathbf{M}_{sp} * \mathbf{M}_{sp}}.
\label{eqn:consth}
\end{equation}
We use this as our fiducial choice for cell size. It can be made even more conservative by scaling $h$ up by some factor, or by choosing $h=\lambda_{\max}$, where $\lambda_{\max}^2$ is the eigenvalue of the $n_{\max}$ principal component.

Whatever our choice for $h$, we can compute $N_{\mathrm{cells}}$ when using $n$ principal components in the model as
\begin{equation}
N_{\mathrm{cells}}(n) = \frac{\prod_{i=1}^n S \lambda_i}{\prod_{i=1}^n h_i}.
\label{eqn:Ncells}
\end{equation}

An example of calculating $N_{\mathrm{cells}}(n)$ for a temperature slice of the red clump using Equation~\eqref{eqn:consth} is shown in Figure~\ref{fig:rcNcell}. 


\subsection{Jackknife}
\label{sec:jackknife}

To determine the uncertainty in our results, we employed a jackknife technique to verify the robustness of the number of principal components we measure. For an individual slice of the red-clump or red-giant stars in $\teff$ or $\log g$, we divided the stars randomly into 25 bins and computed the number of principal components for 25 subsamples created by using star from all but one of the bins. We then found the mean number of components $\bar{n}_{\mathrm{max}}$ across the 25 subsamples and computed the variance in the average with
\begin{equation}
	\sigma^2_{\mathrm{max}} = \frac{J-1}{J}\sum_{j=0}^J \left(n_{\mathrm{max},j} -\overline{n_{\mathrm{max}}}\right)^2,
\end{equation}
where $J$ is the total number of subsamples and $n_{\mathrm{max},j}$ is the number of principal components as computed from the $j$'th subsample. 

\begin{figure*}
\centering
\includegraphics[width = \linewidth]{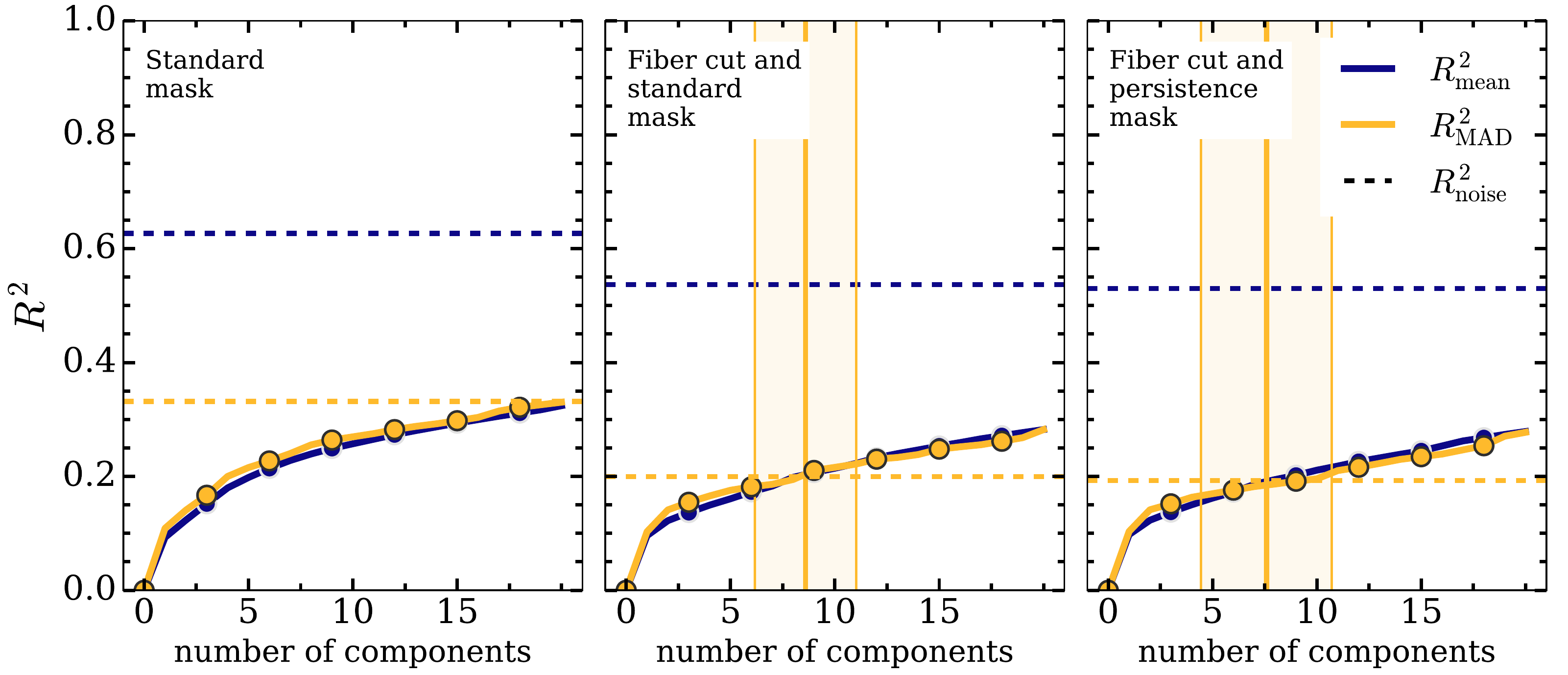}
\caption{This figure has the same structure as Figure~\ref{fig:n6819R2}, but was created using data from a 200 K slice of the red clump sample ($4700\,\mathrm{K} < \teff < 4900\,\mathrm{K}$). We applied a jackknife technique to this temperature slice, dividing it into 25 subsamples (see \S\ref{sec:jackknife}). Computing $\Rsq$ curves for these subsamples gives a mean number of principal components needed to explain the subsample (solid vertical line) and the range of 68\% confidence on this number (shaded region).}
\label{fig:R2RC}
\end{figure*}

An example of jackknife results is shown in Figure~\ref{fig:R2RC}. The bold vertical line marks the location of the average number of principal components computed for 25 subsamples $\bar{n}_{\mathrm{max}}$, while the shaded regions show the uncertainty $\sigma_{\mathrm{max}}$ on this number.

\section{Results}

\label{sec:app}

We apply the algorithm described in the previous section to analyze the chemical space represented by several slices in the stellar properties of our main samples of red-giant and red-clump stars. We first test that our method reproduces the expected behaviour in open and globular clusters.

\subsection{Cluster samples}

\label{sec:OC}

We use open clusters for which APOGEE has observed at least 10 stars: NGC 6819 (30 stars), M67 (24 stars), and NGC 2158 (10 stars). Our primary test subject was NGC 6819, as it had the most stars observed and had the fewest stars observed with persistence-flagged fibers. The $\Rsq$ curve for this cluster is displayed for several masking techniques in Figure~\ref{fig:n6819R2}. For each masking choice, two $\Rsq$ curves are shown, one for the mean-based variance and one for the MAD-based variance. The $\Rsq$ curves do not differ much, especially in the case of aggressive masking, but the $\Rnoise$ values shown by the dashed horizontal lines more than doubles when using mean-based variance instead of MAD. $\Rnoise$ is defined as $\Rnoise = 1-V_{\mathrm{noise}}/V_{\mathrm{data}}$. $V_{\mathrm{noise}}$ is always computed in the same way, but $V_{\mathrm{data}}$ changes with choice of mean-based or MAD-based variance, and is larger when mean-based variance is chosen, indicating that the distribution of flux variations is non-Gaussian with heavy tails. The mean-based variance is very sensitive to outlying stars, when we would prefer to sample the most densely populated part of spectral space. For this reason, we use the MAD-based variance to compute all $\Rsq$ statistics from the other samples. We also choose the most aggressive masking technique to reduce the influence of instrumental effects in finding principal components. While this approach works well for NGC 6819, it masks all but four of the stars in M67 and completely masks NGC 2158. However, much like for NGC 6819, we find that we require zero principal components to model these clusters even with only a standard mask. In truly homogeneous open clusters, we expect the number of components to be zero, since all stars in such a cluster would be tightly localized in spectral space, with scatter attributable only to measurement uncertainty. Reproducing this expected result with our analysis is confirmation that our algorithm behaves as expected.

We performed an additional test on a sample of 71 stars from the globular cluster M13. Applying our most aggressive mask reduces the sample to 18 stars. Using MAD-based variances yields two principal components needed to model this cluster. A brief analysis of these principal components by investigating absorption lines tabulated in \citet{Smith2013} reveals that the first component has a strong signal at the only unmasked aluminum line in our spectrum ($1.675515\,\mu\mathrm{m}$). This identifies that line as a location where spectra vary strongly, and indicates that Al is an important element for distinguishing between the stars. M13 stars have been measured to have significant spread in [Al/Fe] of about 2.5 dex when their traditional abundances are computed \citep{Meszaros2015}, with a smaller spread of about 0.5 dex in [Mg/Fe]. We also see signal in our first principal component around some Mg lines, but it is weaker than that displayed in the Al line. Both signals are stronger than those typically seen at wavelengths corresponding to absorption features for other elements, and the Al feature is much stronger than the local mean. However, there are also sharp absorption-like features in the principal components that do not correspond to absorption features for these two elements, and neither element is particularly strong in the second principal component. We describe how we might go beyond this preliminary analysis in \S\ref{sec:discussion}, but the confirmation of significant spread in Al is further evidence for the success of our approach to analyzing chemical space.

\begin{figure}
\centering
\includegraphics[width = \linewidth]{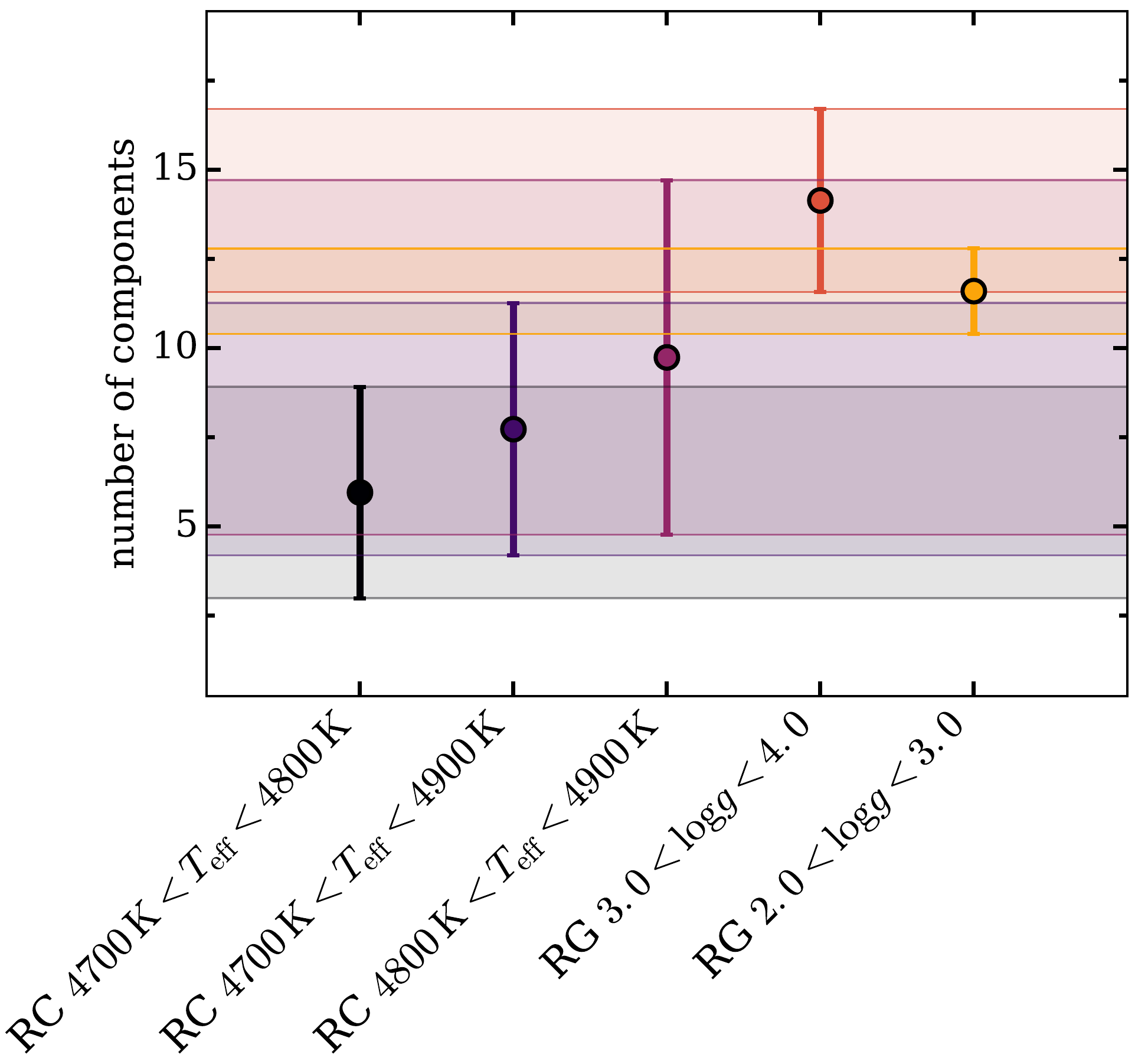}
\caption{Comparison of the number of principal components needed to explain the variations in the spectra above the measurement noise for red clump (RC) and red giant (RG) subsamples. The errorbars are derived from jackknife analysis of 25 subsamples for each slice, and their range is shaded to facilitate comparison between samples.}
\label{fig:samplesnvec}
\end{figure}

\begin{figure}
\centering
\includegraphics[width = \linewidth]{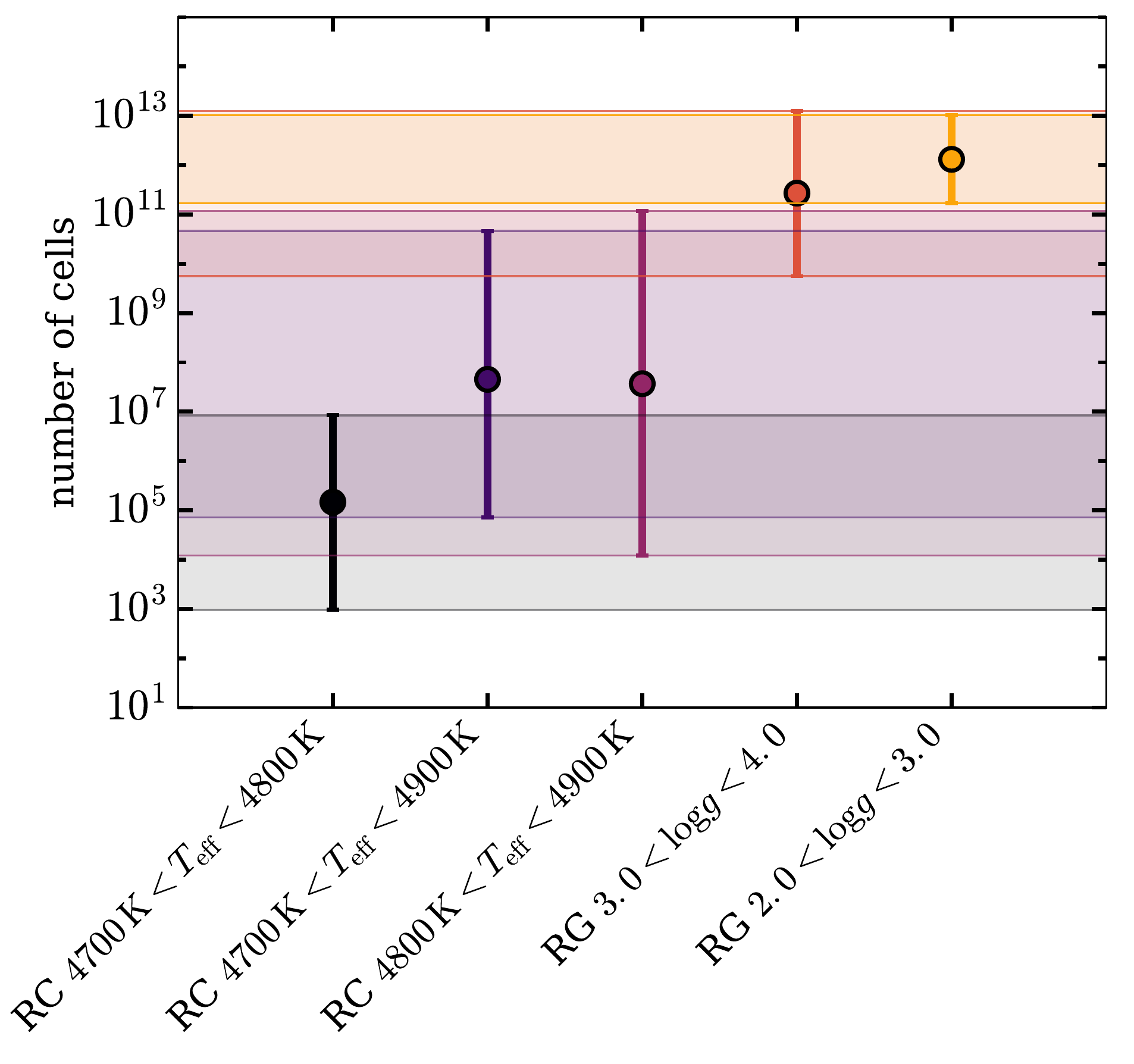}
\caption{Comparison of the number of chemical space cells spanned by each sample, where error bars show the possible range of $N_{\mathrm{cells}}$ according to the range of principal components represented by the points in Figure~\ref{fig:samplesnvec}. These values for $N_{\mathrm{cells}}$ were computed in the optimistic case where $h_i \rightarrow h$ using Equation~\eqref{eqn:consth} in Equation~\eqref{eqn:Ncells}}
\label{fig:samplesncells}
\end{figure}

\subsection{Red clump and red giant samples}
\label{sec:RC}

With successful tests on the cluster samples, we apply the algorithm to the larger red clump and red giant samples, now assessing not only the dimensionality of chemical space (the number of principal components), but also the granularity in that space (the number of chemical space cells). 

\subsubsection{Principal components}

We show an example $\Rsq$ curve using a sample drawn from the peak of the red clump distribution in Figure~\ref{fig:R2RC}; this sample is hereafter referred to as the example slice ($4700\,\mathrm{K} < \teff < 4900\,\mathrm{K}$).  The intersection between $\Rsq$ and $\Rnoise$ gives approximately 8 principal components needed to model this slice, although this varies slightly with different masking approaches. As we explain in \S\ref{sec:OC}, we choose the most aggressive mask (cutting out stars observed with persistence-affected fibers and masking persistence affected pixels), and use an MAD-based approach to computing the variance in the sample. This corresponds to the rightmost panel of Figure~\ref{fig:R2RC}. Using the same choices for mask and variance computation, we compare the example slice with several other samples: two sub-slices of the example slice with $4700\,\mathrm{K} < \teff < 4800\,\mathrm{K}$ (3191 stars), and $4800\,\mathrm{K} < \teff < 4900\,\mathrm{K}$ (2547 stars), and two slices of the red-giant stars with $2\, < \log g < \, 3$ (6532 stars), and $3\, < \log g < \, 4$ (3618 stars). This choice of red-giant slices covers a large fraction of the red-giant stars, even after our cuts on fiber, and overlaps with the peak of the red clump distribution. These samples are compared in Figure~\ref{fig:samplesnvec}, where we show the number of principal components needed to explain each sample above the level of measurement noise. Errorbars are computed through the jackknife technique described in \S\ref{sec:jackknife}. 

Figure~\ref{fig:samplesnvec} reveals that the red-clump slices and the red-giant slices are internally consistent, and that the red-clump and red-giant slices are mostly consistent with each other, with overlap occurring around 10 principal components. The red-giant samples display a slightly higher dimensionality. In general, the samples with more stars tend to have smaller errorbars, pointing to a consistent dimensionality regardless of which stars are used. Although the uncertainties derived from the jackknife approach are not insignificant, generally of order 2-4 components, it is worth noting just how markedly consistent all the samples are. Even when performing very different cuts on stellar properties, the underlying distribution being sampled has similar dimensionality, reflective of the fact that we constrain stellar evolutionary phase and sample a similar region of the Galaxy with each slice.

\subsubsection{Chemical space cells}
A variation of a few in the number of principal components can translate into a large change in the number of chemical space cells $N_{\mathrm{cells}}$ spanned by a sample. An example calculation of $N_{\mathrm{cells}}$ in the fiducial case of constant cell size (Equation~\eqref{eqn:consth}) is shown for our example slice in Figure~\ref{fig:rcNcell}. It is clear that for this constant cell size ($h\approx 0.007$), $N_{\mathrm{cells}}$ is a steep function of the number of principal components; this holds for all of our slices and results in a range of 10 orders of magnitude in the number of cells across the uncertainty in $n_{\max}$ for all of the five slices, shown in Figure~\ref{fig:samplesncells}. Modelling the $N_{\mathrm{cells}}$ for each slice as a power law in the number of principal components yields the following relation when median parameters are taken across the five slices $N_{\mathrm{cells}} \approx 10^{9\pm2} \times (5\pm 2)^{n-10}$. This is strongly determined by our choice to assume constant cell size approximately given by the measurements uncertainties, and is very sensitive to that cell size. Scaling the cell size by a factor of $k>1$ reduces $N_{\mathrm{cells}}$ by a factor of $k^n$, with $n$ the number of principal components. Letting $k=5$ brings $h$ to $\approx 0.035$ and $N_{\mathrm{cells}}$ to about 50 at $n=10$, with an increase by a factor of just 1.1 with each additional principal component. We can make a similarly conservative choice for constant $h$ by letting it be $\lambda_{\max}$, the squareroot of eigenvalue corresponding to the $n_{\max}$ principal component for that sample. For this choice of $h$, $N_{\mathrm{cells}}$ is approximately 1000 with $n=10$, increasing by a factor of about 1.3 with each additional principal component. 

It is worth noting that implicit in this analysis is that we have already fit out [Fe/H] for each slice. For our samples, [Fe/H] has a MAD-based deviation of about 0.4 dex, with a measurement uncertainty of about $\lesssim 0.05$ dex. This corresponds to a factor of $\sim$10 more cells for the values shown in Figure~\ref{fig:samplesncells}. When this additional spread is factored into $N_{\mathrm{cells}}$, we find that the total number of chemical space cells inhabited by our sample is 
\begin{equation}
	N_{\mathrm{cells,\,total}} \approx 10^{10\pm2} \times (5\pm 2)^{n-10}
\end{equation}

These values for $N_{\mathrm{cells,\,total}}$ may seem high, but they are less dramatic when considered in a high dimensional space. In such a space, the chemical space cells are distributed so that each axis is divided into $\left(N_{\mathrm{cells,\,total}}\right)^{1/n}$ bins within which stars cannot be distinguished. For the fiducial choice for $h$ there are approximately 10  such bins per axis. For more conservative choices of $h$, this number dips to just over 1 bin per axis. This further reinforces the need for small measurement uncertainties if we are to distinguish stars in this high dimensional space.

\section{Discussion}
\label{sec:discussion}

The work presented here has very promising implications for performing chemical tagging in spectral space, an approach so far unexplored by other chemical tagging studies. These studies have enjoyed some success, particularly the blind abundance space recovery of known clusters in APOGEE data by \citet{Hogg2016}. However, this success in the strong limit of chemical tagging runs counter to the results of numerous other experiments. Recent work by \citet{Blanco-Cuaresma2015} combined abundances from several surveys to show significant overlap between the chemical signatures of open clusters, which persists even after a larger number of abundances are considered \citep{Blanco-Cuaresma2016}. An earlier blind tagging experiment by \citet{Mitschang2014} also combined several surveys and also found that there is likely significant overlap between birth clusters. This view is further supported by the high rate of abundance space \textit{doppelganger} reported in the APOGEE data by \citet{Ness2017}. These results are much more aligned with theoretical predictions like \citet{Ting2015a}, who used a model of star formation and migration to show that overlap between birth cluster chemical signatures would make identification of individual birth clusters quite challenging. This outlook can be improved by increasing the survey size, but the precision of the abundance measurements is also crucial.

Each of these chemical tagging experiments, regardless of the success they report, relies on model derived abundances, and assumes a corresponding uncertainty in these abundances based on uncertainties in the measured spectra and the model used to reproduce them. The uncertainty in chemical abundances is typically $\lesssim 0.1$ dex for current surveys (e.g., \citealt{Smiljanic2014}, \citealt{Holtzman2015}). The magnitude of this uncertainty has a strong influence on the possibility of distinguishing clusters of stars in abundance space. The increased precision we find by using spectra directly provides greater confidence in our discriminatory power in chemical space, although this is not without its own limitations.

\begin{figure*}
\centering
\includegraphics[width = \linewidth]{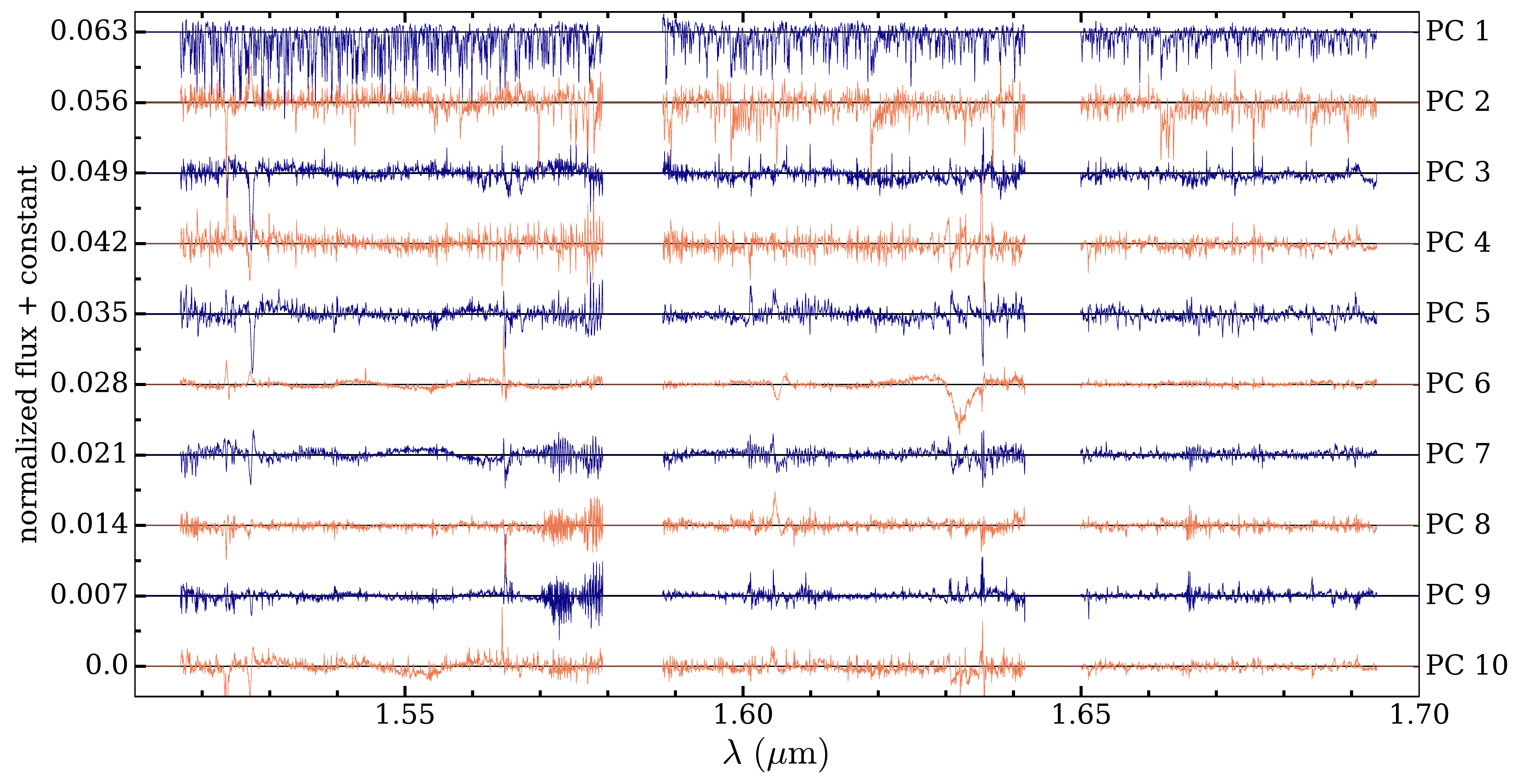}
\caption{The first 10 principal components derived from red-clump stars with $4700\,\mathrm{K}\,<\,\teff\,<\,4900\,\mathrm{K}$, scaled by the median coefficient used to model the spectra in the sample. The PCs are ranked by their importance in explaining the variance in the data, with 1 the most important and 10 the least important. The components have been arbitrarily separated by 0.007 in normalized flux. The first four principal components are dominated by narrow features, with large scale structure first making a noticeable appearance in blue side of principal component 5. There are also broad features around 1.635 $\mu$m in component 6 and 1.605 $\mu$m in components 6, 7, and 8. The gaps from $\sim$1.58 to $\sim$1.59 $\mu$m and $\sim$1.64 to $\sim$1.65 $\mu$m are the spaces between the APOGEE detectors.}
\label{fig:eigvec}
\end{figure*}

\begin{figure*}
\centering
\includegraphics[width = \linewidth]{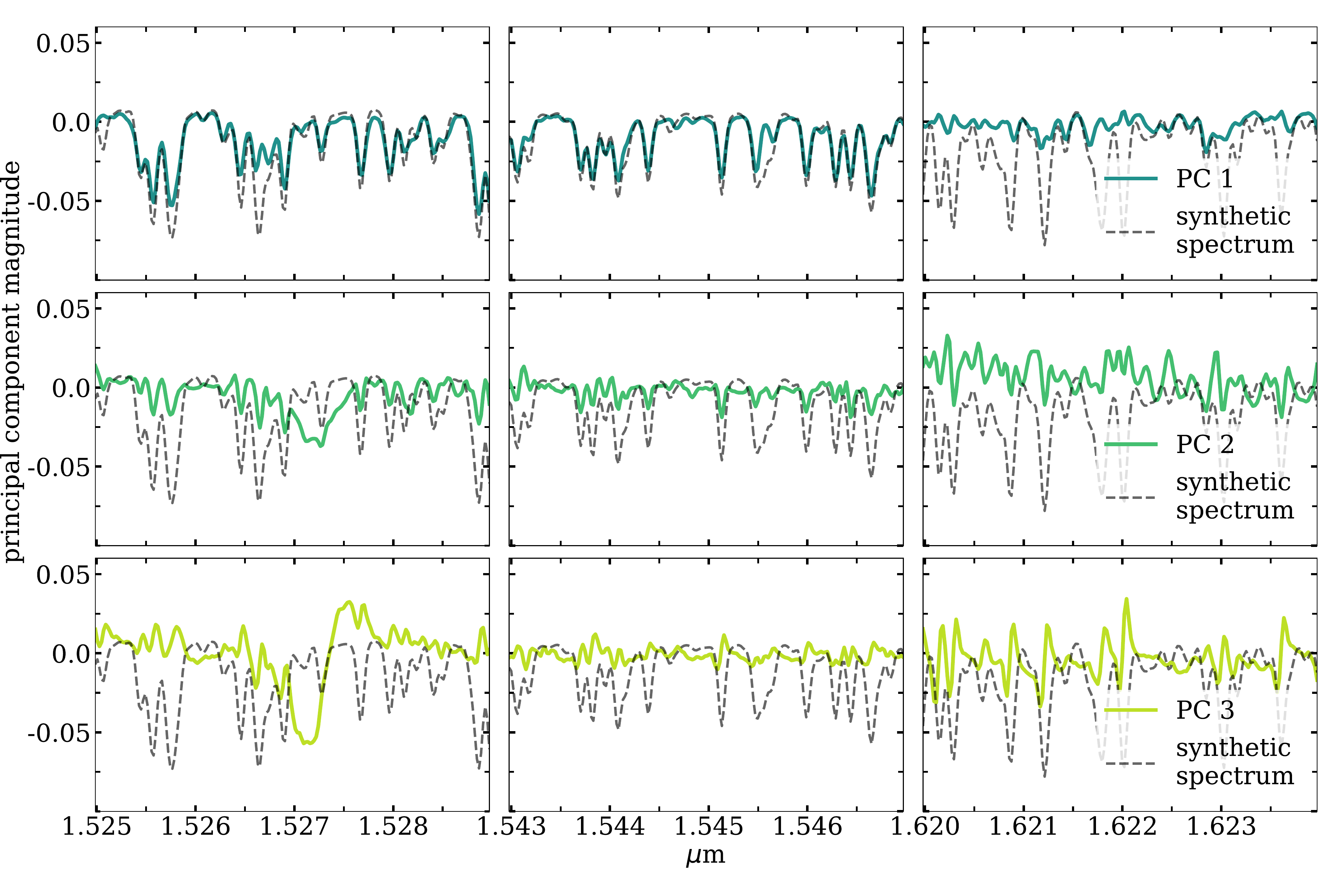}
\caption{This figure shows a series of zoomed in regions on the first three principal components (solid lines). These are compared with a synthetic H-band spectrum (dashed lines) with $T_{\rm eff}$=4809 K, $\log g$ = 2.62 and abundances set to the median of the example slice. The synthetic spectrum has been arbitrarily scaled to roughly match the magnitude of the principal components for the sake of comparing the locations of features. The first two columns show locations where features in the principal components match the locations of features in the synthetic spectrum, while the last column shows a wavelength range where the synthetic spectrum is strong but the principal components do not track that strength}
\label{fig:zoomeigvec}
\end{figure*}

\begin{figure*}
\centering
\includegraphics[width = \linewidth]{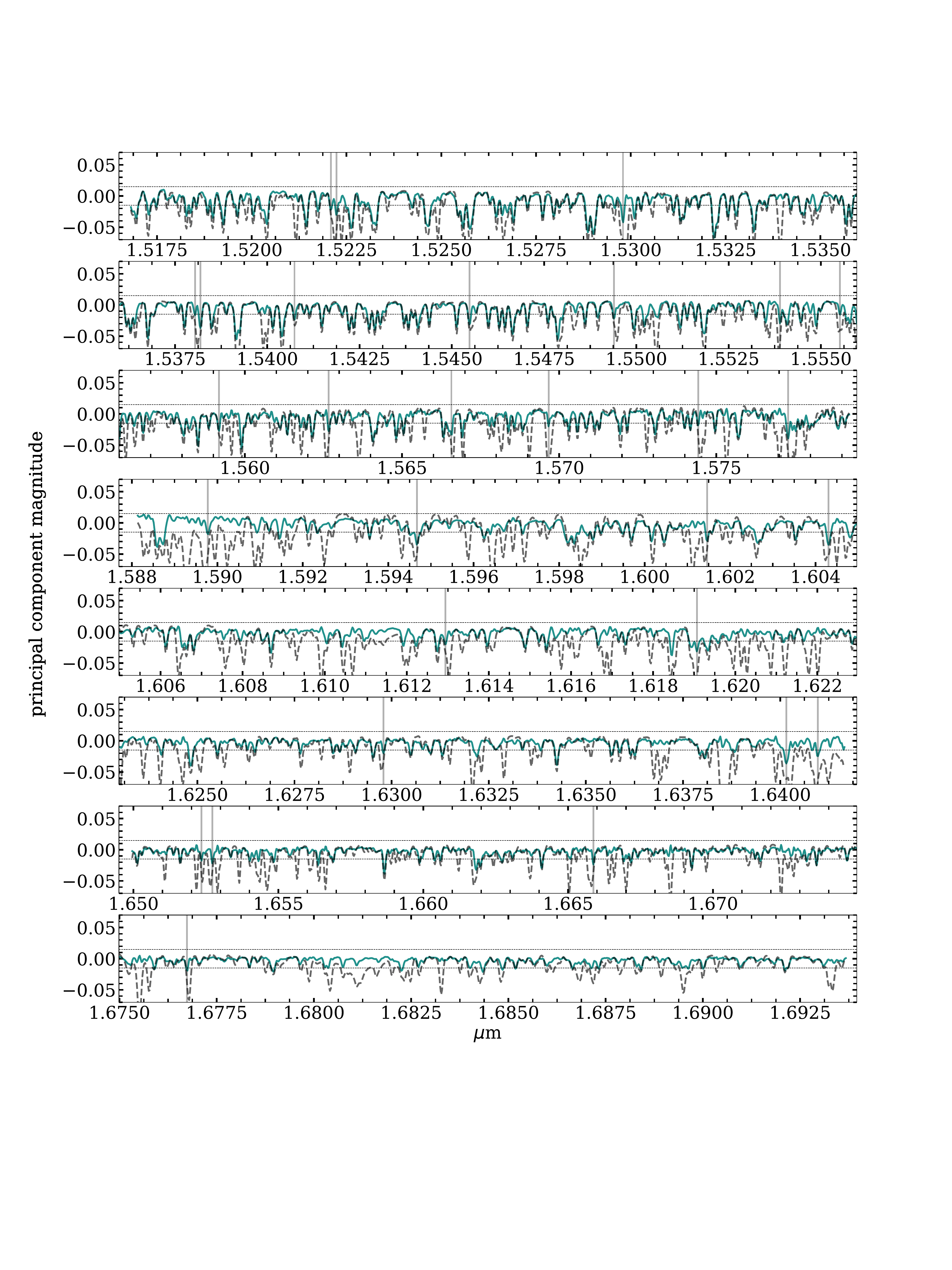}
\caption{The first (and most important) principal component needed to explain variance in the example slice (solid blue line), compared to a rescaled synthetic spectrum with the median properties of that slice (grey dashed line). A simple root-finding algorithm was used to identify peak locations in a re-scaled synthetic spectrum and in the principal component where peak height exceeded a threshold value (horizontal dashed lines). Where no synthetic spectrum match was found to a peak in the first PC, a grey vertical line marks its location. Overall, 84\% of the peaks in the first PC had matching peak locations in the synthetic spectrum within 1.5 pixels.}
\label{fig:zoomPC1}
\end{figure*}


The high number of relevant principal components we find is an upper limit on the dimensionality of the space accessible in the $H$-band, and as such is not inconsistent with previous predictions by \citet{Ting2015a} of about 4-5 chemical space dimensions for APOGEE. Our higher bound may indicate that fully using the spectral space allows for a more nuanced approach to strong chemical tagging. However, we must interpret our results with caution, as they rely on a few crucial assumptions. 

\subsection{Critical assessment of assumptions}

\textit{Reliance on ASPCAP:} Despite our general concerns about using model-derived parameters to describe stars, we do rely on the ASPCAP pipeline for radial velocity corrections, visit combination, continuum normalization, and for our fit parameters $\teff$, $\log g$ and [Fe/H], for which we assume the uncertainties are negligible compared to the measurement uncertainties when doing polynomial fits. We have tested the effects of performing our own continuum renormalization using a polynomial model with continuum pixels identified by the Cannon \citep{Ness2015}, but found our results to be nearly identical to those found prior to the renormalization. However, an approach that analyzes spectra at the individual visit level and/or makes use of stellar parameters derived from more accurate techniques like asteroseismology may offer a stronger limit on the number of principal components needed to describe chemical space.

\textit{Non-chemical effects on spectra:} In addition, we have made a few other assumptions as part of our process. We have assumed that $\teff$, $\log g$ and [Fe/H] adequately model non-chemical difference between stars, and we now discuss some possible sources of difference not accounted for by this model. We expect a polynomial in these properties  to be a valid model for the impact of convective mixing or atomic diffusion on surface abundances, since these effects are largely deterministic functions of mass, age and metallicity. However, many stellar properties cannot be described with such a simple model; in particular, rapid rotation (which some red giants have exhibited, see: \citealt{DeMedeiros1999}, \citealt{Massarotti2008} for examples) or strong surface activity would affect the shape of absorption features. We expect general photospheric properties like micro- and macro-turbulence to correlate with the properties we chose to fit, but given that these serve as broad descriptions of a detailed velocity field, nuances in that field would also influence a spectrum on the narrow wavelength scales we would like to associate with chemistry. There are also many effects that would modify spectra in non-trivial ways independent of intrinsic stellar properties. Binary or multiple systems may result in superimposed stellar spectra, or undergo mass transfer scenarios that could modify surface abundances.  Intervening material could also play a role in shaping a spectrum, whether it be incompletely masked atmosphere or diffuse interstellar bands \citep{Zasowski2014}. Instrumental effects also influence the spectra; persistence has already been highlighted as troublesome in the APOGEE detectors, and is sufficiently complex that it is likely not completely removed even by our most aggressive approach to masking it. Changes in line spread function as a function of observing fiber or varying spectral resolution could also introduce small variations in spectral features not due to intrinsic abundance differences. While we make cuts in the fibers we use, the results we have presented do not explicitly account for the variations between fibers. Including the full-width half-maximum of the line spread function associated with each fiber in our polynomial fits does not change the number of principal components above the level of our jackknife-derived confidence regions and overfits our cluster data. A future approach might attempt to account for fiber LSF in individual visit spectra before combining them, but this would not account for other contaminating effects, many of which could produce both chemistry mimicking narrow features as well as large scale structure in the observed spectra.

\textit{Quadratic model for non-chemical effects:} We have also assumed that a quadratic function in $\teff$, $\log g$, and [Fe/H] is a good model for how these properties change based on the results of previous work; it is possible that this over-fits the actual data variations. Our preliminary tests on cluster data give the expected number of principal components, which indicates that the quadratic model is sound. However, if it were over-fitting our polynomial parameters would still be imprinted on the spectrum, and we would expect a relationship between any of the three polynomial parameters and the model coefficients used to create a linear combination of principal components for each star. We tested for such a relationship by computing the correlation coefficient between the model coefficients and each of $\teff$, $\log g$, and [Fe/H], and found that these coefficients were low, especially when compared to the correlation coefficients that result from comparing the model coefficients with the stellar abundances from ASPCAP. Given this and the nature of our principal components shown in Figures~\ref{fig:zoomeigvec} and~\ref{fig:zoomPC1} and discussed below, we feel confident that our quadratic fit does not excessively remove chemical information.

\textit{Neglecting non-linear cross terms in quadratic models:} Our choice to fit quadratically in just three parameters has included neglecting the cross-terms between those parameters and the derived abundances for each star, which might seem invalid when considering that they are necessary in the polynomial spectral models created by the Cannon \citep{Casey2016a}. We have tested the validity of neglected the cross-terms using the polynomial spectral model developed in \citep{Rix2016} to accurately model APOGEE red giant spectra. We compared the spectra created using their full quadratic model with those from a reduced model that had only the $\teff$, $\log g$, and [Fe/H] quadratic terms (removing the non-linear cross terms with other abundances) and found that in the narrow range of parameters that define our sample, the variance introduced by neglecting those cross terms was two orders of magnitude less than the overall variance in our dataset. This small correction to the variance does not significantly change $R^2$ and so makes no difference to the number of principal components derived for each of our slices. While our assumption to neglect the abundance cross-terms holds for the region of parameter space occupied by our sample, future applications of our algorithm would need to account for this assumption.

\textit{Reliance on APOGEE noise model:} Our final assumption was that the noise model provided by APOGEE accurately represents the measurement uncertainties, and that these measurement uncertainties are independent between pixels. It is likely that this is not the case, and future work will look for changes in our result when a full uncertainty covariance matrix is incorporated into both the EMPCA algorithm and the $R^2_{\rm noise}$ statistic.

\subsection{Interpretation of the principal components}

Although our assumptions seem validated given current data by our cluster results and additional tests, some evidence of such large scale structure remains in the principal components we derived from our example slice, shown in Figure~\ref{fig:eigvec}. These components have been scaled by the median coefficient with which they were multiplied to model the data and thus represent their typical contribution to observed spectra. The components have also been ranked in order of their importance for explaining the variance in the data, with the top component (PC1) being the most important. Since these form a basis for the data, the locations and relative magnitudes of features inform us about the importance of particular wavelengths. Features with the same sign change together in spectra while features with opposite signs anti-correlate in spectra. The signs themselves can be folded into coefficients for individual stars and are therefore less important. Without aggressive masking and fiber cuts, the first few principal components were dominated by large scale structure, a strong indicator that persistence was influencing the analysis. However applying more rigorous masking results in the mostly narrow features of Figure~\ref{fig:eigvec} with larger structure beginning to appear in the fifth-most important component. In the sixth, this large scale structure dominates the blue part of the APOGEE detector, and appearing again in the seventh. The sixth component also exhibits are broad feature near $1.635\,\mu$m, while components 6, 7, and 8 show another broad feature near $1.605\,\mu$m. These remaining large scale features betray the presence of residual non-chemical effects in spectra that are not adequately downweighted by measurement uncertainty. We tested multiple methods of continuum removal on the original ASPCAP spectra in an attempt to mitigate these large scale effects, but applying these methods resulted in similar principal components, and larger scale variations were not fully removed. However, these large scale variations are relegated to principal components less important for explaining variance in the data

Despite these residual large scale structures and possible sources of contamination listed above, our results are promising. The first (and therefore most important) principal component in Figure~\ref{fig:eigvec} strongly resembles an absorption spectrum, and the other components also exhibit narrow absorption-like features. At least some of these are not intrinsic chemistry: for example the narrow feature at 1.5273 $\mu$m is a diffuse interstellar band already known in APOGEE \citep{Zasowski2014}. However, to understand the other narrow features we compare them with a synthetic spectrum created with Turbospectrum \citep{Turbospectrum} from an ATLAS9 atmosphere \citep{atlas9} having properties that the median values of the example slice ($\teff = 4809$ K, $\log g = 2.62$, [M/H]=-0.11, [$\alpha$/Fe]=0.06). Figure~\ref{fig:zoomeigvec} shows this comparison for the first three principal components in three different wavelength regions (including one containing the DIB mentioned above). The synthetic spectrum has been arbitrarily rescaled for the sake of comparison, so the y-axis scale is similarly somewhat arbitrary. However it is clear that there is strong agreement between the \textit{locations} of features in the principal components and the locations of absorption lines in the synthetic spectra. The different wavelength regions were chosen to highlight the fact that areas with strong absorption lines in the synthetic spectrum (rightmost column) do not always correspond to significant magnitude in the principal component (although the scales are arbitrary, they are consistent for each line).

To demonstrate the general quality of this agreement, we show the comparison for the first principal component in its entirety in Figure~\ref{fig:zoomPC1}. To quantify the level of agreement, we used a simple root-finding algorithm to identify peaks in the principal component and in the synthetic spectrum, then matched peaks that agreed within 1.5 pixels. Peaks found in the first principal components that had no counterpart in the synthetic spectrum are marked with vertical gray lines in Figure~\ref{fig:zoomPC1}. In total, 84\% of the peaks in the first PC above a magnitude threshold have corresponding absorption features in the synthetic spectrum. Similar analysis on the other principal components reveals 55\% agreement in peak locations for the second PC, with roughly 30\% agreement for subsequent principal components. Supplementing this compelling comparison are the results from our open cluster tests. Finding the expected zero principal components for the open clusters and two principal components for M13 indicates that many of the potential non-chemical effects we listed above must be relatively unimportant for APOGEE data.

\subsection{Directions for future work}

Although we offer some comparisons in Figures~\ref{fig:zoomeigvec} and~\ref{fig:zoomPC1}, it is beyond the scope of this paper to perform the detailed analysis required to fully understand the principal components in Figure~\ref{fig:eigvec}. However, pursuit of this knowledge is worthwhile, as it is possible that these components highlight absorption features from elements not yet identified in APOGEE spectra. Even if this is not the case, it may be useful to use the principal components to find the abundances that are most important for distinguishing stars, facilitating future abundance-based chemical tagging studies.


Other directions of future work would involve streamlining and improving our algorithm.  One primary goal is to further improve the quality of the input spectra. To do this with APOGEE data, a better approach to modelling instrumental effects is needed. One way to remove large-scale effects is to add components $\alpha\,\mathbf{BB}^T$ to the covariance matrix $\mathbf{V}$ in Equation~\eqref{eqn:chisquared} with a large prefactor $\alpha$ and $\mathbf{B}$ a smooth function of the wavelength to be removed. For large $\alpha$, $\textbf{B}$ is then an approximate eigenvector of the data covariance matrix that the EMPCA algorithm will therefore attribute to noise rather than intrinsic variation. By using multiple such components that together describe a large fraction of possible large-scale signals, this approach would effectively remove large-scale wavelength trends from the data. Similarly, we have assumed that pixel-to-pixel correlations within each spectrum are negligible, but using the full pixel covariance version of the EMPCA algorithm, while resulting in slower computations, would take these correlations into account. We may also be able to reduce some of this structure by using spectra before they have been normalized by the ASPCAP, performing our own visit combination and continuum normalization. Slightly simpler to implement would be taking an iterative approach to modelling and subtracting out the large scale structure observed in the principal components, either with median filtering or a more involved  technique like independent component analysis.  

Applying the algorithm to other data sets would sidestep some of the challenges of modelling APOGEE instrumental noise (although this would undoubtedly introduce new sources of contamination). This could also provide access to wavelength regions which may contain features due to elements not represented in the $H$-band. Comparing derived principal components for similar samples in different surveys may help identify structure due to instrument noise or local atmospheric conditions rather than intervening interstellar medium or stellar properties. Increasing sample size would also further populate the chemical space, and aid in cluster finding. The GALAH survey alone plans to observe $\sim 10^6$ stars, increasing from the full APOGEE DR12 data set by almost an order of magnitude. Making full use of these larger samples will require refinement of our algorithm to speed up matrix operations.

Improving our algorithm by relaxing some of our assumptions would likely result in a correspondingly improved set of principal components, but the primary goal of this work was to lay the groundwork for performing chemical tagging by directly using stellar spectra. To that end, we also highlight the possibility of using the principal components found here to test chemical tagging in spectral space. Projecting spectra along these principal components locates those stars in chemical space, after which cluster finding algorithms can be applied to identify overdensities. Successfully recovering open clusters after mixing them with the larger slices will be necessary to determine whether tagged overdensities can be associated with birth clusters. Another way to test whether chemical space over densities can be safely interpreted as birth clusters is to make use of synthetic spectra with well defined properties. Creating synthetic open clusters and attempting to recover them from a larger synthetic sample will place constraints on the chemical properties of birth clusters that can be accurately chemically tagged. This will also help determine the properties of the clusters represented in the APOGEE datasets presented here.

\section{Conclusions}
\label{sec:conclusion}
Using a sample of spectroscopic observations of red-clump and red-giant stars taken with APOGEE, we investigate chemical space using the full spectrum. Our tests on open clusters confirm that these clusters are chemically homogeneous with current measurement precision. We investigate our primary red clump/red giant sample by making slices in the most populated parts of the HR diagram. Using an aggressive mask to mitigate the effects of detector persistence and polynomial fit to remove non-chemical stellar properties, we employ EMPCA to show that about 10 principal components are needed to explain the variance in our chosen datasets. For our fiducial choice of chemical space cell size, we can describe the number of chemical space cells as approximately 
\begin{equation}
	N_{\mathrm{cells,\,total}} \approx 10^{10\pm2} \times (5\pm 2)^{n-10}
\end{equation}
where one factor of 10 is due to the spread in iron abundance removed by our initial polynomial fit, and $n$ is the number of principal components. This quantity could further increase by a factor of $\approx$(SNR/100)$^n$ with improved signal to noise in future surveys. From the simulations of \citet{Ting2015a}, this high number of chemical space cells implies that a high-resolution survey of $\gtrsim 10^6$ stars in the $H$-band could detect a large number of individual birth clusters that are dominated by real members.

 Making a larger and thus more conservative choices for cell size reduces the total number of cells in 10-dimensional space to be of order 100-1000, increasing by a factor just greater than 1 with additional principal components. This swift decrease in $N_{\mathrm{cells}}$ highlights the importance of small uncertainties if using chemical tagging to identify over-densities in chemical space is ever to be successful in the strong limit of finding birth clusters. However the high number of principal components we consistently find across HR diagram slices indicates that stars do populate a high dimensional chemical space, which is promising for distinguishing chemical signatures associated with individual birth clusters. 

\section*{Acknowledgements}

NPJ and JB received support from the Natural Sciences and Engineering Research Council of Canada. JB also received partial support from an Alfred P. Sloan Fellowship and from the Simons Foundation. The authors thank the Flatiron Institute for hospitality during part of the period during which this research was performed.

Funding for the Sloan Digital Sky Survey III has been provided by
the Alfred P. Sloan Foundation, the U.S. Department of Energy Office of
Science, and the Participating Institutions. SDSS-IV acknowledges
support and resources from the Center for High-Performance Computing at
the University of Utah. The SDSS web site is \url{www.sdss.org}.

\bibliographystyle{mnras}
\bibliography{chemical-tagging}

\end{document}